\documentclass[fleqn,10pt]{wlscirep}

\usepackage[utf8]{inputenc}
\usepackage[T1]{fontenc}
\usepackage{amsmath}    
\usepackage{amsfonts,amssymb}
\usepackage{bm}         
\usepackage{graphicx}   

\usepackage{multicol}
\usepackage{pdflscape}
\usepackage{colortbl}
\usepackage{booktabs}   
\usepackage{float}
\usepackage{lineno}
\usepackage{setspace}
\usepackage{tocloft}
\usepackage{datetime2}

\usepackage{caption}
\usepackage{subcaption}

\usepackage{enumitem}
\setlist[enumerate]{parsep=0pt}
\usepackage[none]{hyphenat} 
\usepackage{verbatim}
\usepackage{lipsum}     
\usepackage{kantlipsum} 

\usepackage{orcidlink}

\usepackage[numbers,square]{natbib} 

\newcommand\aap {Astronomy and Astrophysics}

\newcommand\ao {Applied Optics}

\newcommand\apjl {Astrophysical Journal, Letters}              
\newcommand\apjs {Astrophysical Journal, Supplement}

\newcommand\ra { 282.11 $\pm$ 0.05  }  
\newcommand\dec{ 4.56  $\pm$ 0.05 }  
\newcommand\gl{ 36.65 $\pm$ 0.07 $^\circ$ }              
\newcommand\gb{ 2.74  $\pm$ 0.08 $^\circ$ }              
\newcommand\parx{ 0.142 $\pm$ 0.123  }                      
\newcommand\dist{  4185$\pm$ 638 }                                   
\newcommand\pma{  -2.55$\pm$0.83  }                      
\newcommand\pmd{   -5.69$\pm$0.88  }      
\newcommand\fo{ 10.42$\pm$ 0.51  }                         
\newcommand\fb{  14.55$\pm$ 0.13  }                         
\newcommand\rc{  2.69 $\pm$ 0.19   }                       
\newcommand\rt{  23.25$\pm$ 6.88  }

\newcommand\CE{ 2.42  $\pm$ 0.09 }                          
\newcommand\dm{ 12.87$\pm$ 0.21 }                           
\newcommand\distph{ 3746 $\pm$ 432 }                     
\newcommand\age{ 125.0$\pm$ 12.30 }                          
\newcommand\Zini{0.0153}            
\newcommand\MH{0.0151}    

\newcommand\Ns{ 1740$\pm$ 29.83 }   
\newcommand\Vr{ 41.6 $\pm$ 7.1 } 
\newcommand\Vt{ 131$\pm$ 33.4 } 
\newcommand\tha{  -114.8$\pm$ 7.54$^\circ$  }   
\newcommand\Nrv{66     }   
\def\footnote#1{\textcolor{blue}{ \textbf{ [#1] }}}
\title{Discovery of a new open cluster as a companion to Czernik 38 cluster and its associated complex tide , using Gaia DR3}
\author[1,*]{Nasser M. Ahmed\,\orcidlink{0000-0002-7701-3032}}

\affil[1]{National Research Institute of Astronomy and Geophysics (NRIAG), 11421 Helwan, Cairo, Egypt}

\affil[*]{Nasser M. Ahmed \textbf{:} nasser\_ahnmed@yahoo.com}
\begin{abstract}
Our earlier accepted research emphasized a detailed investigation of the Czernik 38 cluster through the utilization of Gaia DR3 data. We  have discovered a new star cluster positioned 32 arcmin from its center, with coordinates $\alpha$=\ra  and $\delta$=\dec.  
Our suspicion is that this clump might form a new cluster \textbf{(hereafter referred to as Nasser 1)} and could be a companion to Czernik 38 in a binary cluster system. \emph{In this ongoing research, we aim to investigate this matter in depth and derive the physical parameters of the Nasser 1 cluster using Gaia data, in comparison to Czernik 38.}~ 
We employed the pyUPMASK Python package with HDBSCAN algorithm to analyze membership of the Nasser 1 cluster to get its physical parameters. Our innovative method involves establishing a membership probability threshold value for each radius, as opposed to utilizing a singular value for the entire cluster. 
At each radius, we calculate the probability threshold value that yields a number of stars equivalent to that inferred from the King model.
 The foremost conclusion of our research is that this clump of stars is actually a new open star cluster.
This newly identified cluster possesses a clearly defined CMD characterized by an age of \age Myr, a distance modulus of \dm~ mag, and a significant  color excess of \CE, respectively. In addition, it exhibits a distinct King profile.  Further examination reveals that both clusters possess the same age, distance, proper motion, and radial velocity, even same high reddening value, within acceptable error limits.  This implies that they together constitute a fantastic young primordial binary system characterized by a substantial reddening, located at the periphery of the Carina-Sagittarius spiral arm. 
Both clusters are affected by the impacts of differential rotation tides, as evidenced by their elongations in the direction of orbital motion. Furthermore, the Nasser 1 cluster is affected by an additional tide, specifically a gravitational tide caused by the Carina-Sagittarius spiral arm.
In addition, the Gaussian mass function indicates that they were formerly an one cluster that has been violently turn apart by differential rotation during its orbital trajectory. 
\end{abstract}
\begin{document}
\label{firstpage}
\maketitle
%
\tableofcontents
%
%
%
%
\section{Introduction}
Open Clusters (OCs) are stellar systems that are homogeneous or nearly so, consisting of several dozen to a few thousand stars with different masses situated at comparable distances. Where, they were originated from the collapse of the same dense molecular clouds.  Consequently, they share identical ages, formed physical conditions, kinematics, and chemical compositions, all while orbiting under the gravitational influence of the galaxy.  Thus, they are significant not only for the study of stellar evolution and as physical laboratories, but they also serve as important subjects for comprehending the structure, kinematics, and characteristics of the Milky Way \cite{Madore2022,Carraro2007, Cantat-Gaudin2020, Lada2003}. \\

Our study, which has been accepted for publication 
\cite{Nasser2025c}, involved a comprehensive analysis of the Czernik 38 cluster utilizing Gaia DR3. In the course of this analysis, we  accidentally identified aclump of stars situated 32 arcmin away from the center of Czernik 38,  at $\alpha=$\ra and $\delta=$\dec. In this research, we will analyze this clump  thoroughly to determine whether it is actually an open cluster, using  CMD, RDP, and proper motion  derived  from  Gaia DR3 data.  Additionally, we will compare these parameters with those of Czernik 38 to ascertain whether they constitute a binary cluster or not.

\textcolor{red}{\textbf{Primordial binary clusters}} or multiple system may form during the gravitational collapse of massive giant molecular clouds and it dense into many gas clumps , with collapsing these clumps forming clusters of similar ages and very close to each other \cite{Darma2021,Arnold2017,Mora2019}. 
Projected separations for the majority of the cluster pairs are less than 20 pc, which also has relatively young ages, and their small age differences between their constituent parts imply that they originated as primordial binary star clusters \cite{Dieball2002}. Their  color-magnitude diagrams (CMDs) is the same (same stars populations). In some cases, perhaps some force might rip apart or tear a rich cluster into smaller clumpy pieces with the same ages, chemical composition, kinematics and distances. Mostly, these clumpy pieces are often elongated.\\

\textcolor{red}{\textit{On the contrary,}} occasionally a cluster pair or system may be formed by tidal capture during a close encounter between them, typically of different ages \cite{Bergh1996} and \cite{Fuente2009}. Then the CMDs for the merger products are a composite of different stars populations \cite{Leon1999}. \\

Before the launch of Gaia, we summarize the observations and research efforts regarding the incidence of binary open clusters in the Milky Way and the Magellanic Clouds.~\textbf{:-}
\begin{enumerate}[topsep=.3em, itemsep=.3em, , labelsep=1.0em, itemindent=0em]
\item [\textbf{1)}]
\textcolor{red}{\textbf{Observationas in Magellanic Clouds,}}  \cite{Rozhavskii1976} initially claimed that the fraction of multiple systems among open clusters in the Milky Way was roughly 20\%. This neglected result was obtained before any research was undertaken to assess the importance of cluster binarity in the Magellanic Clouds or in Milky Way. The first such studies was when \cite{Bhatia1988} published a list of 69 star cluster pairs in the Large Magellanic Cloud (LMC) with projected separations of less than 18 pc, then it was a topic of interest. Later, \cite{Pavlovskaya1989} proposed the existence of five possible cluster groups. \cite{Hatzidimitriou1990} inferred that the fraction of star clusters pairs in the Small Magellanic Cloud (SMC) is was nearly 10\%. Moreover,  \cite{Loktin1997} provided a catalog of 31 probable multiple systems by restricting the spatial vicinities and age coincidences.  \cite{Fuente2009} argued that the real fraction is  12\%, similar to that in the LMC. These results were later confirmed by an extensive and more rigorous work completed by \cite{Dieball2002}. 
\item [\textbf{2)}] 
\textcolor{red}{\textbf{Conversely, within our Galaxy,}} the only widely recognized binary or double cluster system is the h + $\chi$ Persei pair (NGC 869/NGC 884). However, the true physical distance separating the members of this pair exceeds 200 pc (refer to \cite{Fuente2009}).
However,  according to \cite{Subramaniam1995} , 16 pairs of clusters were identified with a spatial separation of under 20 parsecs, leading to the conclusion that around 8\% of open clusters might actually be binaries, which poses a challenge to the established viewpoint. Despite these findings, it remains underscored (see, e.g., \cite{Dieball1998a}; \cite{Dieball1998b} and \cite{Bekki2004})  that the quantity of cluster pairs within the Milky Way is limited in comparison to that found in the LMC and SMC. This is often interpreted as evidence supporting the higher efficiency of formation of bound stellar groups in those regions. In all these studies, projected distances, rather than three-dimensional distances, were applied.
\end{enumerate}

\textcolor{red}{\textbf{Nowadays in the Gaia era}} where this data base with different release provides full astronomical data (positions, parallaxes, and proper motions) and multi-band photometry for about billions of stars. The existence of binary star clusters in nearby galaxies is a common topic. The high-quality astrometric and photometric data from Gaia DR3 Releases \cite{GaiaCollaboration2023}, combined with new highly efﬁcient tools such as Scikit-learn~\footnote{https://scikit-learn.org/stable/}\cite{Pedregosa2011}  and pyUPMASK\footnote{https://github.com/msolpera/pyUPMASK} \cite{Pera2021} have significantly raised the total number of open clusters found. Then, with the number of known open clusters increasing rapidly and the classification of cluster member stars based on Gaia data becoming rather accurate, there have been fervent searches for binary or multiple systems of open star clusters out there. As example,  \cite{Song2022} identify 14 pairs of open clusters. \cite{Conrad2017} detected 19 groupings, including 14 pairs. \cite{Soubiran2019} provided 21 cluster pairs differing by less than 100 pc in distance and 5 km/s in velocity. Moreover,  \cite{Liu2019} found 56 candidates for star cluster groups using Gaia DR2 data based only on the 3D positions of the star clusters.  \cite{Palma2025} examined the  clusters found in the catalog of 	\cite{Hunt2023,hunt2024improving} and reported the identification of 617 paired systems. \\

The paper is organized as follows. Section \ref{sec:data} outlines the criteria used to extract the initial data sample from Gaia DR3. The cluster's structure, along with its radial density profile, is described in Section \ref{sec:RDP}. In Section \ref{sec:membs}, we present the astrometric analysis, cluster membership determination, and the identification of the cluster's center. The photometric properties of the cluster members  are discussed in Sections \ref{sec:phot}. Furthermore, in section \ref{sec:dynamics_morphology}, the morphology and gravitational tides of the two clusters are discussed. 
Finally, the main conclusions are summarized in Section \ref{sec:summary}. 
%
\section{The Data} \label{sec:data}
In this research, we make use of the Gaia DR3 datasets to analyze the newly discovered cluster. These datasets serve as a reliable foundation for pinpointing cluster members, determining astrophysical parameters, and exploring the cluster's structure and kinematics. Below, we outline the main features and importance of these datasets for the ongoing study.\\

We obtained the data from the Gaia DR3 catalog \cite{GaiaCollaboration2023}. The dataset encompasses sky positions (~$\alpha$, $\delta$), proper motions (~$\mu_{\alpha}\cos\delta$, $\mu_{\delta}$~), and parallaxes, with a limiting magnitude of $G = 21$~mag. Gaia DR3 offers astrophysical parameters for a wide array of celestial objects, which are derived from parallaxes, broad-band photometry, and mean radial velocity spectra. The catalog includes $G$ magnitudes for approximately 1.806 billion sources, $G_{BP}$ magnitudes for close to 1.542 billion sources, and $G_{RP}$ magnitudes for around 1.555 billion sources.\\

\begin{figure*}
\centering
\subfloat[The surface number density of the Nasser 1 cluster, within 50 arcmin.]{\label{fig:surf_mapa}\includegraphics[width=.54\linewidth]{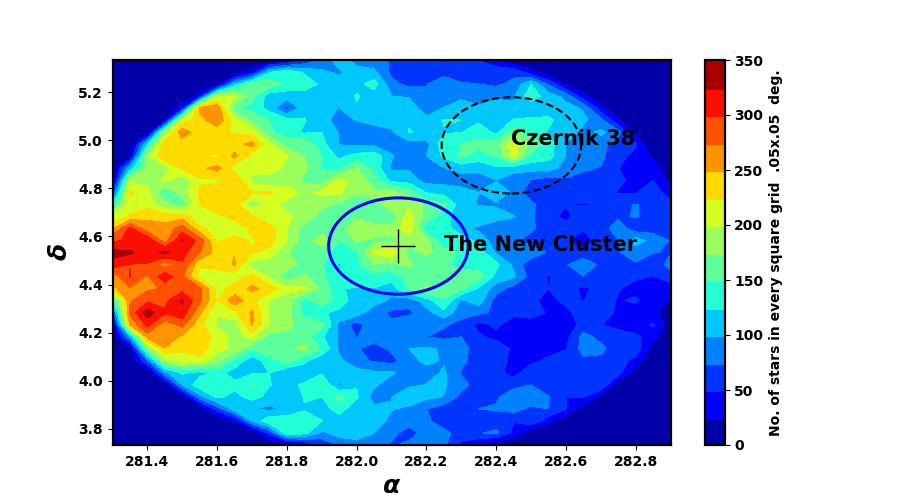}}\hfill
\subfloat[The surface number density of 2 degrees field of view which indicates Cze 38 cluster is on the borders of Carina-Sagittarius spiral arms. The white circle could be a new cluster.  ]{\label{fig:surf_mapb}\includegraphics[width=.43\linewidth]{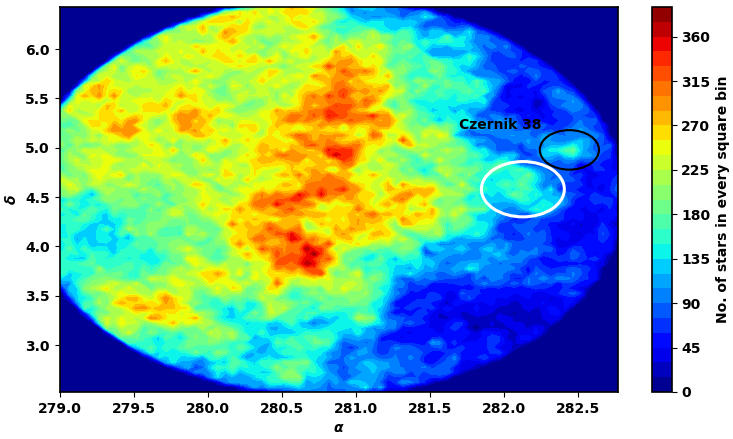}}\hfill
\caption{The surface number density of Czernik 38 using the data of Gaia DR3.}
\label{fig:surf_map}
\end{figure*}

\begin{figure*} 
   \centering
   \includegraphics[width=12.0cm, angle=0]{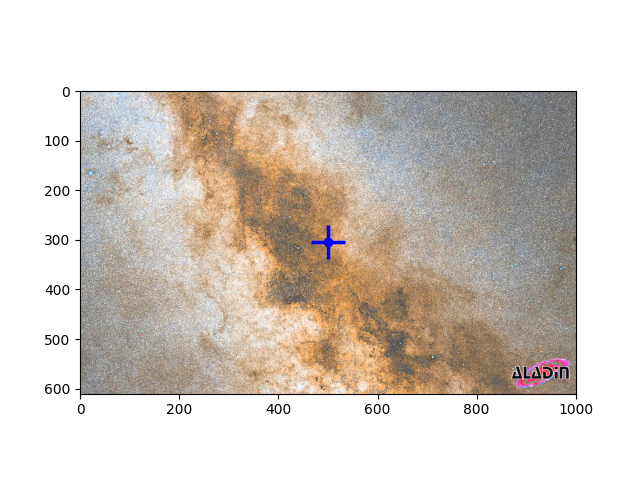}
   \caption{The new cluster's position (blue "+" symbol) is superimposed on the AlaDin (PanSTARS) image depicting the Carina-Sagittarius spiral arm. Take note of the dim area, which is composed of dust and dense gaseous. } 
   \label{fig:sat_arm}
   \end{figure*}

The surface number density is employed as a means to demonstrate the structural characteristics of clusters and their neighboring regions.  As illustrated in Fig.~\ref{fig:surf_map}, the surface number density in the area of the newly identified cluster is obtained from Gaia DR3. Fig.~\ref{fig:surf_mapa} provides a close up view of the Nasser 1 cluster alongside Czernik 38, whereas Fig.~\ref{fig:surf_mapb} presents the surface number density across a two degree field of view, revealing that both clusters are positioned at the edge of the Carina-Sagittarius spiral arms. Additionally, in Fig.~\ref{fig:sat_arm}, the position of the Nasser 1 cluster, marked by the '+' symbol, is superimposed on the AlaDin (PanSTARS) image, which demonstrates its location in relation to the Carina-Sagittarius spiral arm.
%
%
%
\section{The Cluster Radial Density Profile (RDP) } \label{sec:RDP}
To construct the RDP of this cluster, the observed area was divided into concentric rings. The number of stars within each ring, represented as $N_i$, was counted, and the star density was determined by equation $n_i = N_i / A_i$, where $A_i$ denotes the area of the $i$-th ring ($\pi (R_{i+1}^2 - R_i^2)$). Here, $R_i$ and $R_{i+1}$ indicate the inner and outer radii of each ring, respectively, with the definition of the ring radius being:
\begin{equation}  
   r_i = \frac{R_{i} + R_{i+1}}{2}. \label{eq:radius}
\end{equation}
The stars density function $n_{t}(r)$, which represents the total stellar density (including field stars and cluster members), is expressed as:
\begin{equation}
   n_{t}(r) = n_{bg} + n_{c}(r), \label{eq:king1}
\end{equation}
where the background star density is $n_{bg}$ and the cluster member stars density is $n_{c}(r)$.\\  One of the famous profile for cluster stars density that was created by King \cite{King1962} is as follows:
\begin{equation}
   n_t(r) = 
   \begin{cases}  
      n_{bg} + k \left[ \dfrac{1}{\sqrt{1+(r/r_c)^2}} - \dfrac{1}{\sqrt{1+(r_{cl}/r_c)^2}} \right]^{\beta}, & r \leq r_{cl}, \\
      n_{bg}, & r > r_{cl},
   \end{cases} 
\label{eq:king62}   
\end{equation}
where $r_c$,  $n_o$ and $r_t$ represent the core radius, central density, and cluster radius, respectively. And $k$ is:
\begin{equation}
   k = n_o \left[ 1 - \dfrac{1}{\sqrt{1+(r_{cl}/r_c)^2}} \right]^{-\beta}
\end{equation}
and 
\begin{equation*}
   n_c(r_c) = k \left[ \dfrac{1}{\sqrt{2}} - \dfrac{1}{\sqrt{1+(r_{cl}/r_c)^2}} \right]^{\beta}
\end{equation*}

We have introduced the $\beta$ index in place of 2. We fit the RDP of the cluster with the  equation \ref{eq:king62}, allowing $\beta$ to take on only the values 1 or 2.  We fit the RDP of the cluster with the equation \ref{eq:king62} , allowing $\beta$ to take on only the values 1 or 2.\\
In this research, the optimal fit occurs at $\beta$ equal to 1; however, this is not universally applicable, as the fitting value is affected by the steepness of the density profile of the cluster. In some cases, $\beta$ can equal 2, leading to a better fit.\\
The stellar density $n_{c}(r_i)$ in the $i$-th ring can be used to estimate the number of member stars in that ring:
\begin{equation}
   N_{cl, i} = n_{c}(r_i) \cdot A_i,
\end{equation}
By summing the cluster density profile up to $r_{cl}$, we can estimate the total number of cluster members inside this radius:
\begin{equation}
   N_{cl} = N_t \;-\; \pi r_{cl}^2 \; n_{bg} 
   \label{eq:starsn}
\end{equation}
In this context, $N_t$ refers to the complete number of stars (comprising members and field stars), while the term $N_{cl}$ indicates the total number of cluster members that can considerably affect the probability cutoff value, as will be examined in section \ref{sec:cutoff}. \\

\begin{figure*} 
   \centering
   \includegraphics[width=.8\linewidth, angle=0]{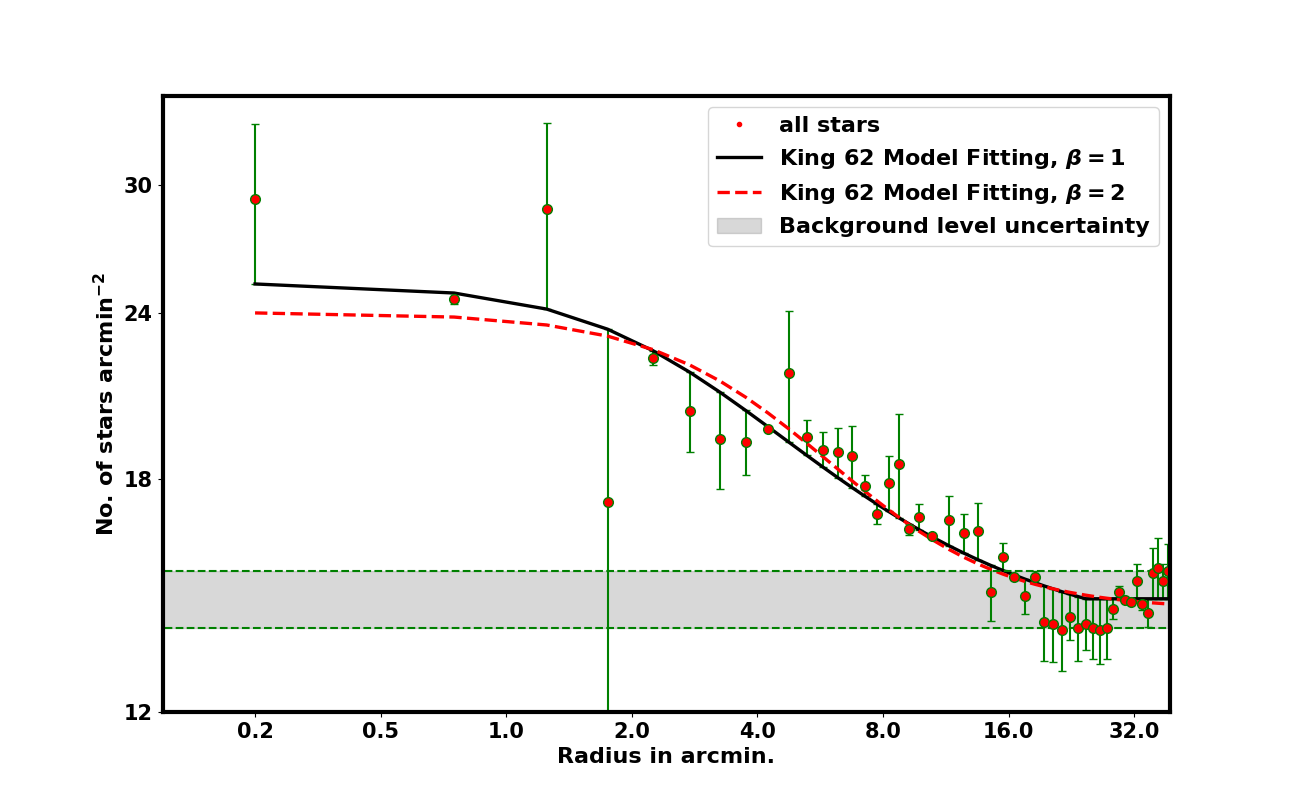}
   \caption{The radial density profile (RDP) of the Nasser 1 cluster. The solid  and dashed  lines represent the King model fits with $B=1$ and $B=2$, respectively. The horizontal strip  indicates the background level uncertainty. \textbf{This figure confirms the existence of stars overdensity in this Nasser 1 cluster.} } 
   \label{fig:rdp}
   \end{figure*}

The structural parameters of the cluster were determined by fitting the King model to the RDP. The background density, $n_{bg}$, was found to be \fb stars~arcmin$^{-2}$ (indicated by the blue dashed line in Fig.~\ref{fig:rdp}). The central density, core radius, and cluster radius were determined to be \fo stars arcmin$^{-2}$, \rc arcmin, and \rt arcmin, respectively (see Table~\ref{tab:king}). Furthermore, the total number of member stars $N_{cl}$ were estimated as \Ns~stars.  The uncertainties in the fitted parameters were estimated using the covariance matrix obtained from the \texttt{curve\_fit} function in the \texttt{scipy} package\footnote{\url{https://scipy.org/}}.
\begin{table*}
\centering
\caption{King model fit parameters. }
\begin{tabular}{| c|c | c | c | c|c |} 
 \hline
The Name &  $n_o$ &  $n_{bg}$ & $r_c$ & $r_{cl}$ & $N_{cl}$ \\ 
& No. $arcmin^{-2}$ & No. $arcmin^{-2}$ & arcmin & arcmin & no of stars\\
 \hline
The Nasser 1 cluster & \fo & \fb & \rc & \rt &\Ns \\
\hline
Czernik 38 & 22.33± 2.24  & 11.6± 0.03   & 1.19 ± 0.02 & 14.36± 7.63 &   938 $\pm$ 61 \\
\hline
\end{tabular}
\label{tab:king}
\end{table*}
%
%
\section{The Membership} \label{sec:membs}
The evaluation of fundamental parameters for star clusters is commonly complicated by the contamination caused by field stars. Historically, the determination of cluster membership relied on photometric and kinematic data.  Nonetheless, the introduction of astrometric data from the Gaia survey has greatly enhanced the precision of kinematic techniques for determining membership. Proper motion and parallax measurements are especially useful in differentiating field stars from those belonging to clusters, given that stars within a cluster generally exhibit comparable kinematic characteristics and distances \cite{Rangwal2019}. In this study, we utilized Gaia DR3 proper motion and parallax data to distinguish between cluster members and nonmembers.
\subsection{HDBSCAN Algorithm}
We employed the Unsupervised Photometric Membership Assignment in Stellar Clusters (UPMASK) algorithm, as developed by \cite{Krone-Martins2014}. This method is non-parametric and unsupervised, eliminating the necessity for prior selection of field stars. An improved version, available in the form of the \textit{pyUPMASK} Python package\footnote{https://github.com/msolpera/pyUPMASK} \cite{Pera2021}, expands the foundational algorithm by including multiple clustering approaches from the scikit-learn library \cite{Pedregosa2011} \footnote{https://scikit-learn.org/stable/}. This library offers over a dozen distinct clustering techniques for unlabeled data, all of which can be utilized in pyUPMASK, including KMS, OPTICS, Mini Batch K-means (MBK), Gaussian Mixture Models (GMM), and Hierarchical Density-Based Spatial Clustering of Applications with Noise (HDBSCAN); their references can be found in \cite{Pera2021}. This feature enables a more adaptable analysis of unlabeled data.

In this study, we employed the HDBSCAN algorithm as described by \cite{Campello2013}, which has been implemented in Python by \cite{McInnes2017}. The HDBSCAN algorithm is considered one of the quickest existing clustering algorithms and represents an advancement over both DBSCAN and OPTICS. It is important to note that DBSCAN functions on the assumption that the criteria for clustering, particularly the density requirement, remains consistent throughout the entire dataset. As a result, DBSCAN may struggle to accurately identify clusters that display differing densities.
HDBSCAN overcomes this limitation by relaxing the assumption of uniformity and exploring a variety of density levels through the creation of an alternative representation of the clustering problem. Moreover, it primarily utilizes a k-means clustering algorithm, which is a technique for classifying data based on its proximity to specified center points. Additionally, it effectively identifies and eliminates noise within a dataset. Consequently, HDBSCAN is recognized as one of the most commonly used and referenced clustering algorithms.

\begin{figure} 
   \centering
\includegraphics[width=0.45\linewidth, angle=0]{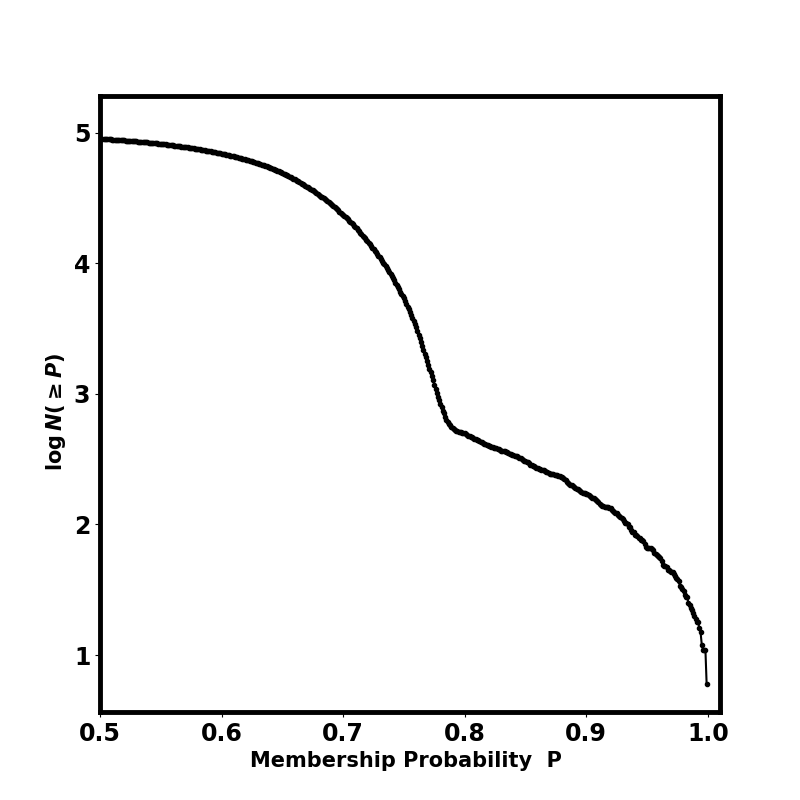}
   \caption{The number of stars as function of membership probability, the output of \textit{pyUPMask} code.} 
   \label{Fig:prob}
\end{figure}

In this research, we employed the 	\textit{pyUPMASK} package alongside the HDBSCAN algorithm to determine the membership probabilities of stars located within the cluster. Gaia DR3 data ($\alpha$, $\delta$, $\mu_{\alpha} \cos \delta$, $\mu_{\delta}$ and $\varpi$) for approximately 234,446 stars within a $50^\prime$ radius were used as input. Fig.~\ref{Fig:prob} shows the total number of stars, N($\geq$P), as a function of their probability~of~membership~P.
%
%
\subsection{The Problem of Probability Threshold Value}
A membership probability threshold of 50\% is frequently utilized in determining membership; however, this criterion may not always be suitable. The ideal cut-off is contingent upon the method used, along with variables such as the density of the surrounding field and the distance from the star to the cluster center. Additionally, the optimal probability cut-off value can differ from one cluster to another. Consequently, it is essential to rigorously evaluate the selection of probability cut-off, as an incorrect threshold could result in either an overestimation or underestimation of cluster members.

Recent studies have explored various probability cut-off values. For instance, \cite{Tarricq2022} employed the same methodology (HDBSCAN) as we did, with a probability cut-off of 50\%. In contrast, \cite{Zhong2022} and \cite{Gao2020} applied UPMASK with a probability cut-off of P > 70\% and the GMM model with P > 80\%, respectively. Therefore, the choice of the probability cut-off value remains a subject of ongoing debate.

\begin{figure*} 
\begin{center}
   \includegraphics[width=14.0cm, angle=0]{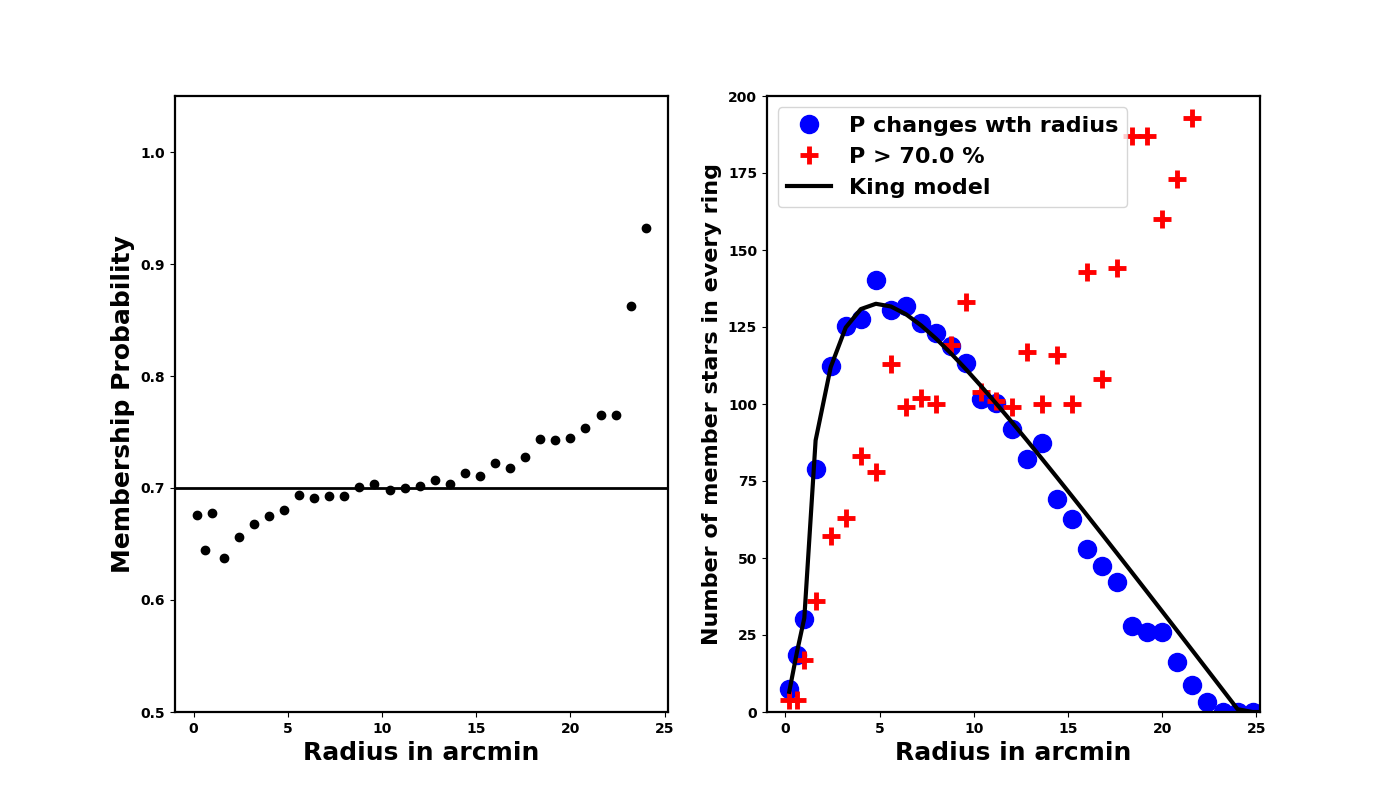}
\end{center}
   \caption{Our approach entails examining the Probability $P_{i}$ at each rang as a function of the radius $r_i$.  In the right panel, the  '+' symbols denote members that have a probability threshold of 70\%, which is entirely inconsistent with the fitting of the King model.} 
   \label{fig:Prob_Radius}
\end{figure*}

In this work, we have applied a novel technique and combine the pyUMASK probabilities with number of stars inferred from fitting King profile fit as follow:
\begin{equation}
   Nm_{i}(P\geq P_i) \;\approx\; n_{c}(r_i) \; A_i
\end{equation}
where $P_i$ is the probability in the $i$-th ring, giving the number of member stars as $Nm_{i}$, which should match the number of stars from the King model, $n_{c}(r_i) \; A_i$, as shown in Fig.~\ref{fig:Prob_Radius}. Furthermore, we have also analyzed the details of this method in \cite{Nasser2025d, Nasser2025c}.
%
%
\section{The cluster kinematics, The Proper motion and  distance } \label{sec:distance}
Open clusters are significant indicators of the evolution of the Galactic disc. The release of Gaia DR3 has enabled the investigation of their kinematics with unprecedented precision and accuracy. The center of the cluster is found at \ra and \dec, which corresponds to the Galactic coordinates l=\gl and b=\gb.

\begin{figure*}
\centering
\includegraphics[width=18cm, angle=0]{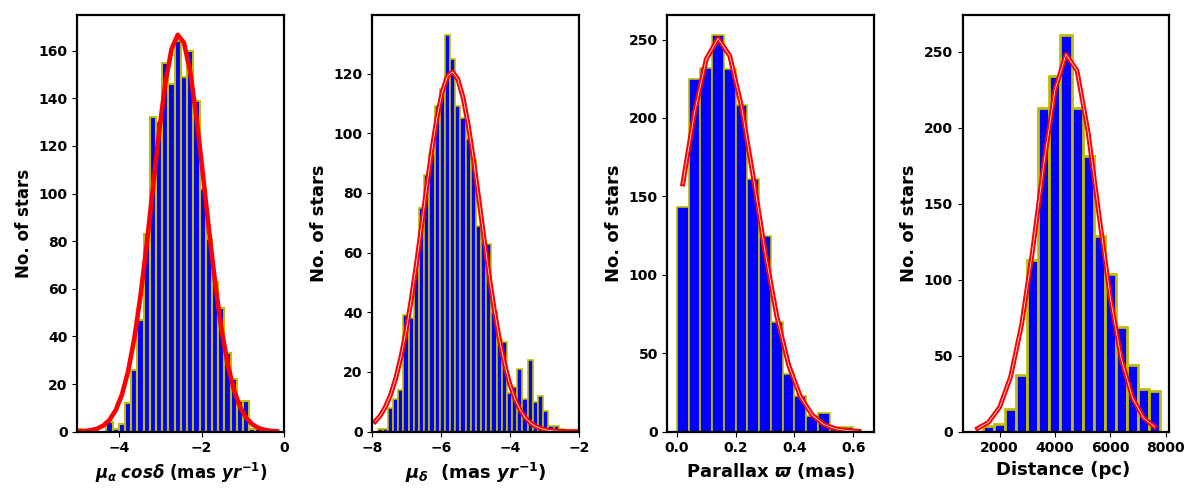}
\caption{The members proper motions, parallaxes and distances histograms. The solid lines are Gaussian fits.}
\label{Fig:param_fit}
\end{figure*}

To improve the precision of the parallax measurements, the parallaxes are modified in accordance with the methodology described in \cite{Lindegren2021}, executed via Python code (gaiadr3\_zeropoint).
The parameters of proper motion and parallaxes are determined using Gaussian distribution fits, as depicted in Fig.~\ref{Fig:param_fit}.  The computed average values are $\mu_{\alpha} \cos \delta$ = \pma mas yr$^{-1}$,  $\mu_{\delta}$ = \pmd mas yr$^{-1}$ and $\varpi=$\parx,  respectively. The detailed results are detailed in Table~\ref{tab:PM}.  Furthermore, Table~\ref{tab:PM} presents a comparison of the parameters values of the new cluster with Czernik 38, indicating that both exhibit nearly identical parameters. \\

Parallaxes ($\varpi$) are vital in the process of distance determination; however, they do not directly equate to distances. This difference is due to the nonlinear relationship that exists between them and the measurement noise that influences distant stars. Even small absolute errors in parallax can result in considerable uncertainties in distance estimations.
Moreover, although parallax may produce negative values, distances themselves cannot be negative.
A potentially more effective strategy could entail the use of a probabilistic method for estimating distances.
\cite{Bailer-Jones2021} introduces a distance catalog that includes \textbf{1.47 billion stars} in Gaia EDR3, which employs a probabilistic technique. Furthermore, we fitted the histogram of these members' distances with a Gaussian distribution. The mean distance to the cluster is determined to be \dist pc, as illustrated in Fig.~\ref{Fig:param_fit}. This value aligns with the findings derived from photometric data within the estimated error margins.\\
\begin{table*}
\centering
\caption{The Coordinates, Proper motion,parallax and distance of the new cluster and Czernik 38 }
\scalebox{0.75}
{\begin{tabular}{| c|c | c || c | c|c || c|c|c|} 
 \hline
 The Name& $\alpha$ & $\delta$ & $\mu_{\alpha}cos\delta$ & $\mu_{\delta}$   & distance & $v_t$& $\theta$ & $V_r $ \\
& deg.  & deg.  & mas yr$^{-1}$  & mas yr$^{-1}$ & pc & km/s. & deg. & km/s.\\
 \hline
The new Cluster & \ra & \dec & \pma & \pmd &\dist & \Vt & \tha &\Vr \\
\hline
Czernik 38& 282.45 ± 0.05& 4.97 ± 0.05 &-2.41 ± 0.328 & -5.263 ± 1.063 & 3666.33 $\pm$ 430.5 & 93$\pm$ 27.16  & -114.33$\pm$ 12.77$^\circ$ & 46.1$\pm$ 8.54  \\
\hline
\end{tabular}
}
\label{tab:PM}
\end{table*}
It is widely recognized that the stars within the cluster exhibit nearly uniform velocities and point towards a specific point.  As a result, the tangential velocities of open clusters, derived from absolute proper motions  (~$\mu = \sqrt{ (\mu_{\alpha}\cos\delta)^2 + (\mu_{\delta})^2}$)  and  distances or parallaxes ($\varpi$),  enable the identification of the type of orbit that the cluster follows. This plays a crucial role in the investigation of the origins and destruction mechanisms of clusters.

The tangential velocity in km/s is given by:
\begin{equation}
v_t = 4.74 \; \mu \; d \; 
\end{equation}

where the constant 4.74 comes from the unit conversion:

\begin{equation*}
\dfrac{(4.84 \times 10^{-6} \; \text{rad}) \; (3.086 \times 10^{13} \; \text{km})}{(3.154 \times 10^{7} \; \text{s})} \approx 4.74
\end{equation*}

\begin{figure}
\centering
\subfloat[The tangential velocity of the new cluster. The solid line is the Gaussian fit with average value \Vt km/s.]{\label{fig:Vt_new}\includegraphics[width=.47\linewidth]{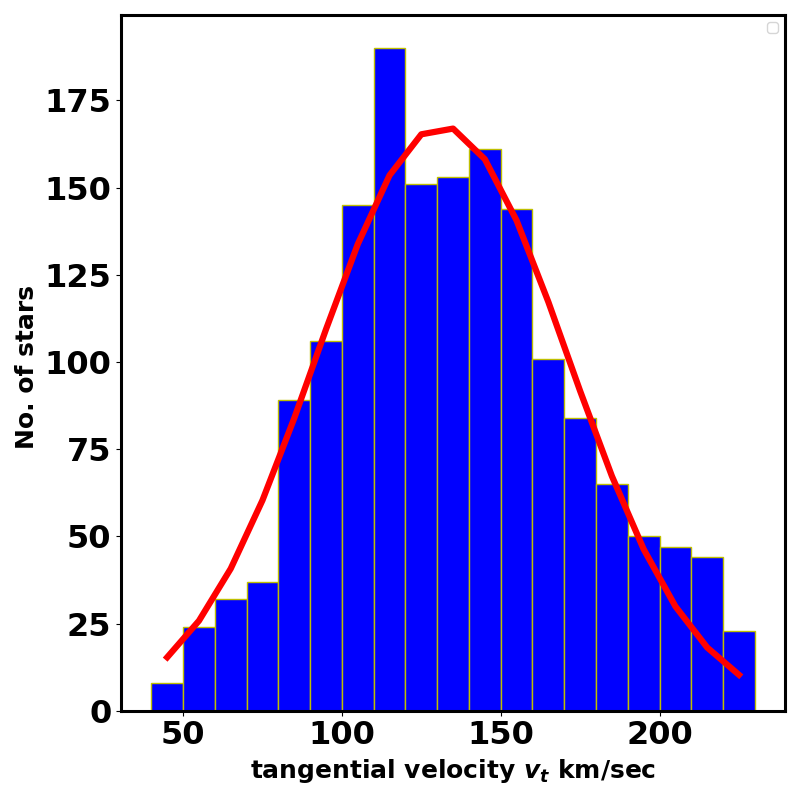}}\hfill
\subfloat[The tangential velocity of the Czernik 38. The solid line is the Gaussian fit with average value 99.6 km/s.]{\label{fig:Vt_cz38}\includegraphics[width=.47\linewidth]{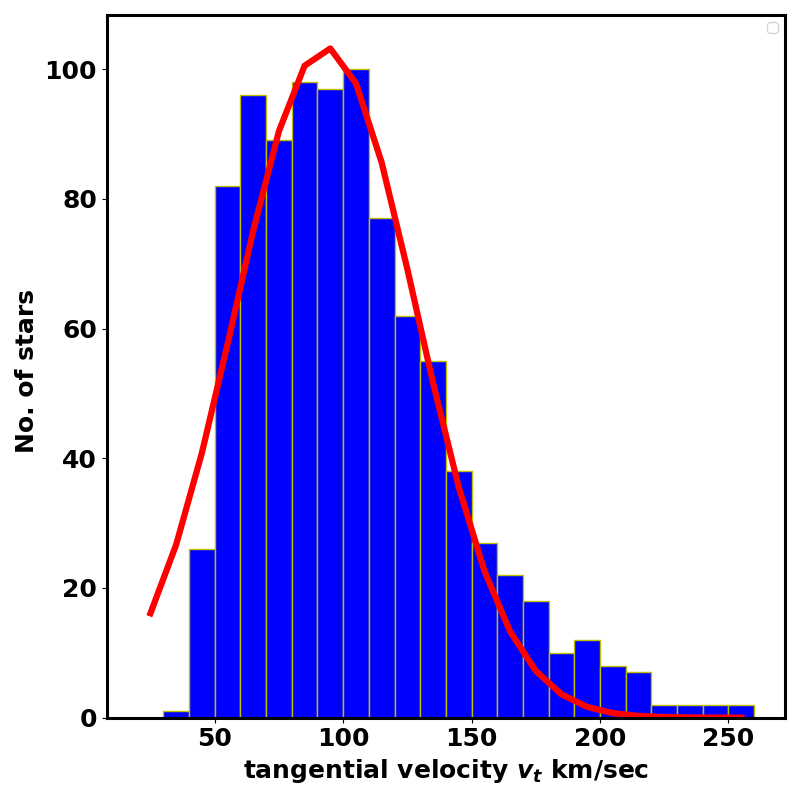}}\hfill
\caption{The average tangential velocities of both the Czernik 38 and the new cluster are identical within the error.} 
\label{fig:Vt}
\end{figure}

Here, $d$ and $\mu$ are the distance, proper motion in parsecs, arcseconds yr$^{-1}$, respectively. Fig.~\ref{fig:Vt_new} shows a histogram of tangential velocity $v_t$ with an average value of \Vt km/s, following a nearly Gaussian distribution.\\

\begin{figure}
\centering
\subfloat[The $\theta$ of the new cluster. The solid line is the Gaussian fit with average value as -112.11 $\pm$ 7.54$^\circ$ ]{\label{fig:theta_new}\includegraphics[width=.48\linewidth]{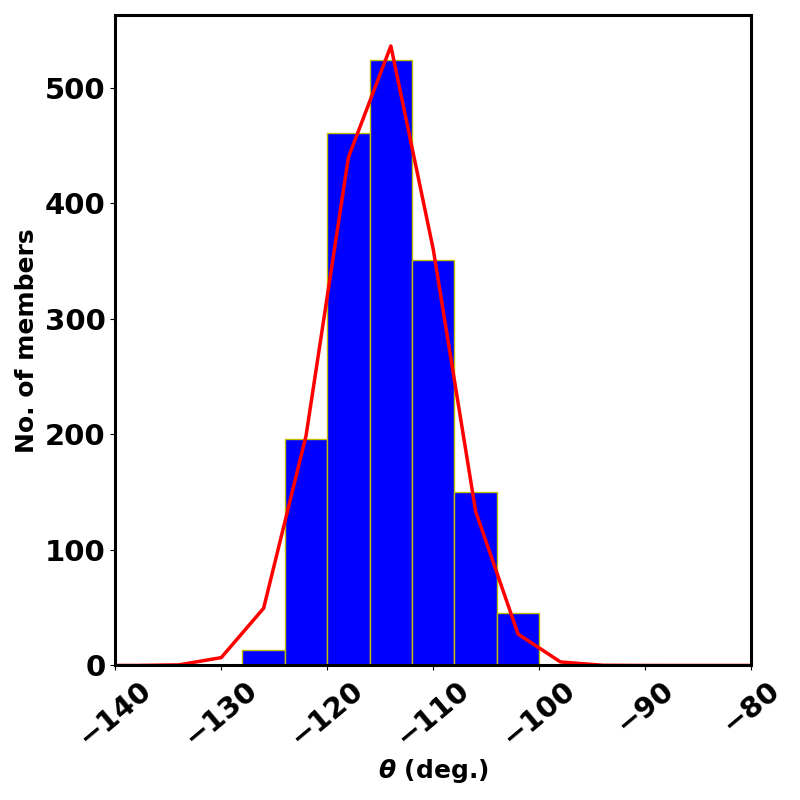}}\hfill
\subfloat[The $\theta$ of the Czernik 38. The solid line is the Gaussian fit with average value  as -109.71 $\pm$ 3.61$^\circ$]{\label{fig:theta_cz38}\includegraphics[width=.48\linewidth]{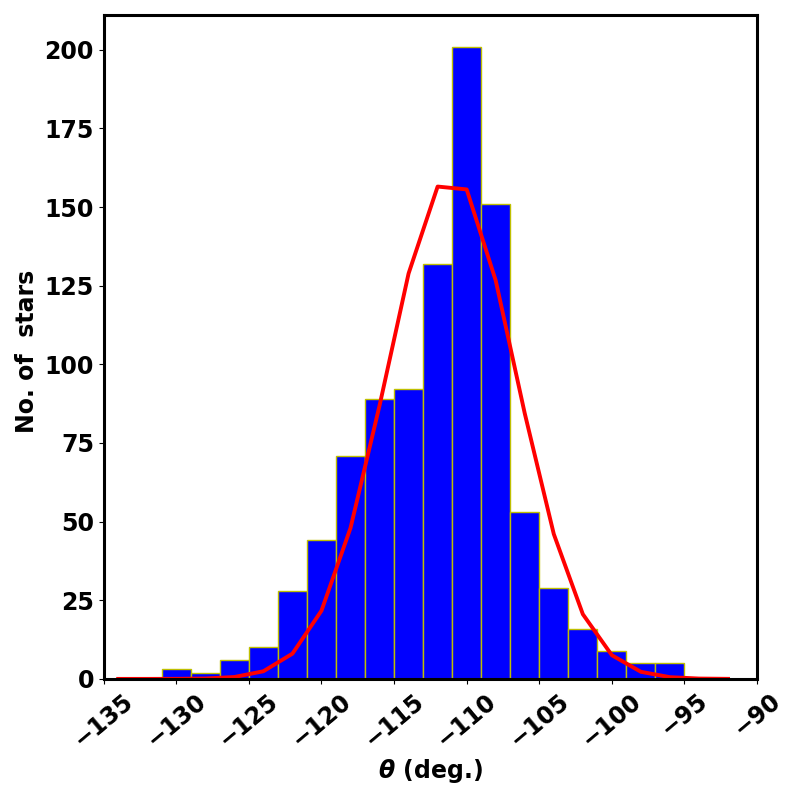}}\hfill
\caption{The new cluster and Czernik 38 are both moving in the same direction towards a unified point.} 
\label{fig:theta}
\end{figure}

As cluster members generally move in nearly the same direction through space, the proper motions of the stars appear to converge at a single point in the sky, referred to as the convergent point. This observed convergence is a perspective effect resulting from the common trajectory of stars in space. Therefore, knowing only the tangential velocity is inadequate; another crucial parameter is the angle $\theta$,  which indicates the direction of the cluster's motion in the $\mu_{\alpha} \cos \delta$ and $\mu_{\delta}$ space.  It is as described by this formula:-
\begin{equation} \label{eq:thet}
\theta = \tan^{-1}\left( \frac{\mu_{\delta}}{\mu_{\alpha} \cos \delta} \right)
\end{equation}
In Figure \ref{fig:theta}, a histogram is displayed showing the $\theta$ values for member stars, with an average angle of \tha. Furthermore, the $\theta$ measurement for Czernik 38 is -114.33$\pm$12.77$^\circ$, revealing that both clusters possess identical values.  Moreover,  we have identified \Nrv~  member stars with radial velocity data with average value \Vr km/s, comparable to the Czernik 38 value of 46.1 $\pm$ 8.54. 
%
%
\section{The Photometry} \label{sec:phot}
The color-magnitude diagrams (CMDs) for open clusters (OCs) make use of empirical isochrones to compare against theoretical models of stellar evolution \cite{Marigo2017, Spada2017}.  CMDs function as powerful tools for determining crucial parameters such as distance, age, and metallicity of a cluster. In addition, by analyzing observed CMDs alongside theoretical isochrones, one can gain valuable insights into the stellar masses within the cluster.  The theoretical isochrones referenced in this study were obtained from the CMD 3.7 website \footnote{http://stev.oapd.inaf.it/cgi-bin/cmd}, employing PARSEC version 1.25s \cite{Bressan2012}.
\subsection{\textbf{Extinction}} \label{sec:extin}
A detailed interstellar dust extinction law is essential for accurately interpreting observations. The extinction coefficients for each passband are influenced by the spectral energy distribution of the source, the interstellar medium, and the extinction itself. Both the color excess ratio (CER), $E(\lambda-\lambda_1)/E(\lambda_2-\lambda_1)$, and the relative extinction, $A_\lambda/A_{\lambda_1}$, are key indicators of the extinction law.

We flow the producer of  \cite{Nasser2025a} and use the method presented in \cite{Wang2019}, we compute the extinction coefficients in the Gaia, 2Mass and BV bands using the relation $A_{\lambda} = a A_V$. For example:
\[
A_G/A_V = 0.789, \quad A_{BP}/A_V = 1.002, \quad A_{RP}/A_V = 0.589
\]
and we get:
\begin{equation}
A_G = 1.88 \times E(G_{BP} - G_{RP})    
\end{equation}
Through isochrone fitting, we derive the color excess and, as a result, the extinction. The intrinsic distance modulus $(m-M)_0$ can be computed using the following equation:
\begin{equation}
\left( m-M \right)_{\text{o}} = \left( m-M \right)_{obs} - A_{\lambda} \label{abs_m}
\end{equation}
where $m$ is the apparent absorbed magnitude, $M$ is the absolute magnitude, and $A_{\lambda}$ is the extinction in the $\lambda$ band. The terms $(m-M)_o$ and $(m-M)_{obs}$ are the intrinsic and observed distance modulus, respectively. Finally, we can drive the isochrone distance $d_{iso}$ as :
\begin{equation}
(m-M)_o \;=\; 5\; \log(d_{iso}) -5,\;\; \\
d_{iso} \;=\; 10^{\dfrac{(m-M)_o +5}{5}} 
\end{equation}
\subsection{ The color-magnitude diagram (CMD) }\label{sec:cmd}
\begin{figure}
\centering
\subfloat[The Gaia CMD of the new cluster with the same isochrone fit of Czernik 38.  The squares are taken from \cite{Abdurro2022}. The stars in rectangle may be white dwarfs.]{\label{fig:CMD_new}\includegraphics[width=.49\linewidth]{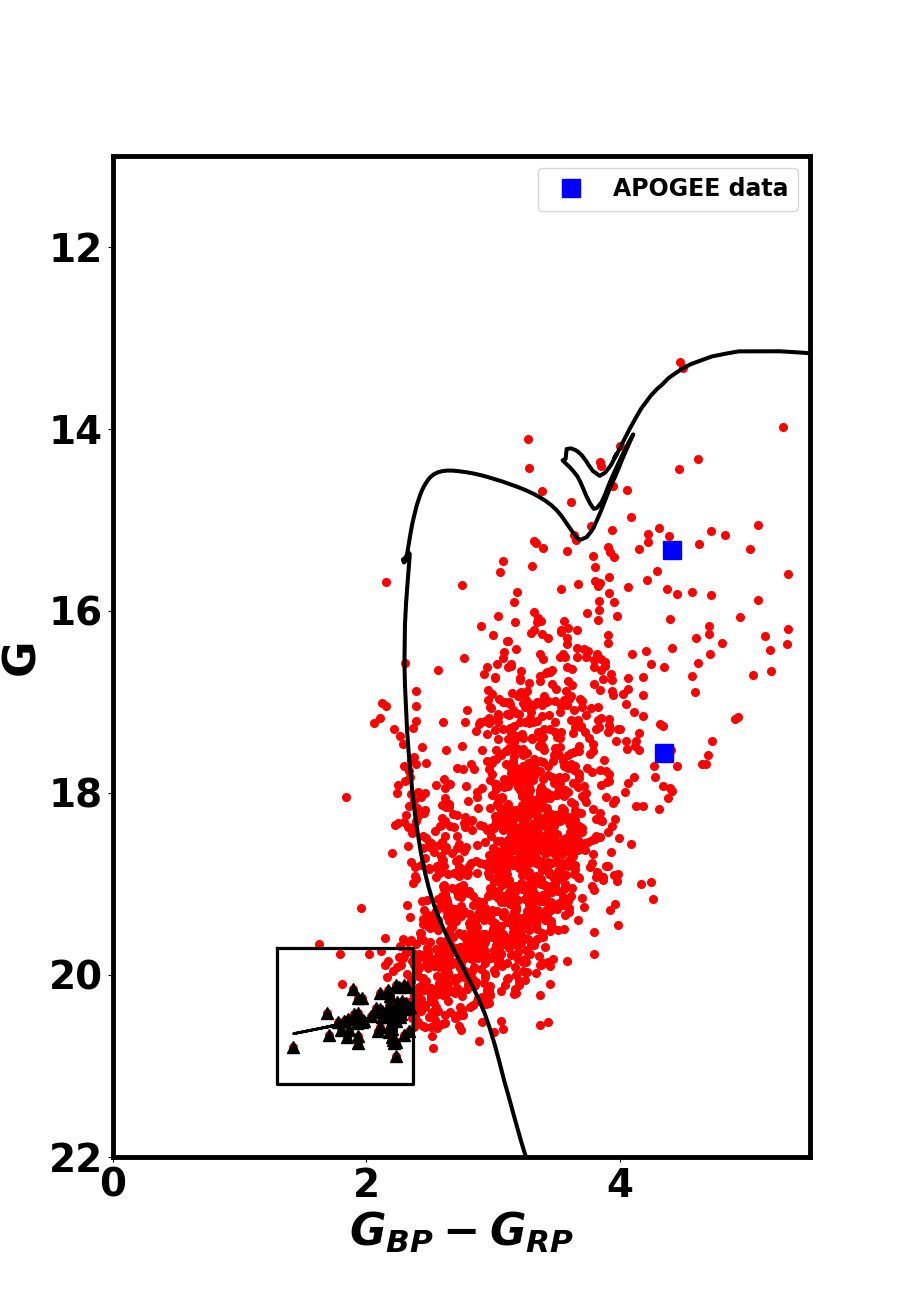}}\hfill
\subfloat[The Gaia CMD of Czernik 38.  The square is taken from \cite{Abdurro2022}. The stars in rectangle may be white dwarfs.]{\label{fig:CMD_cz38}\includegraphics[width=.49\linewidth]{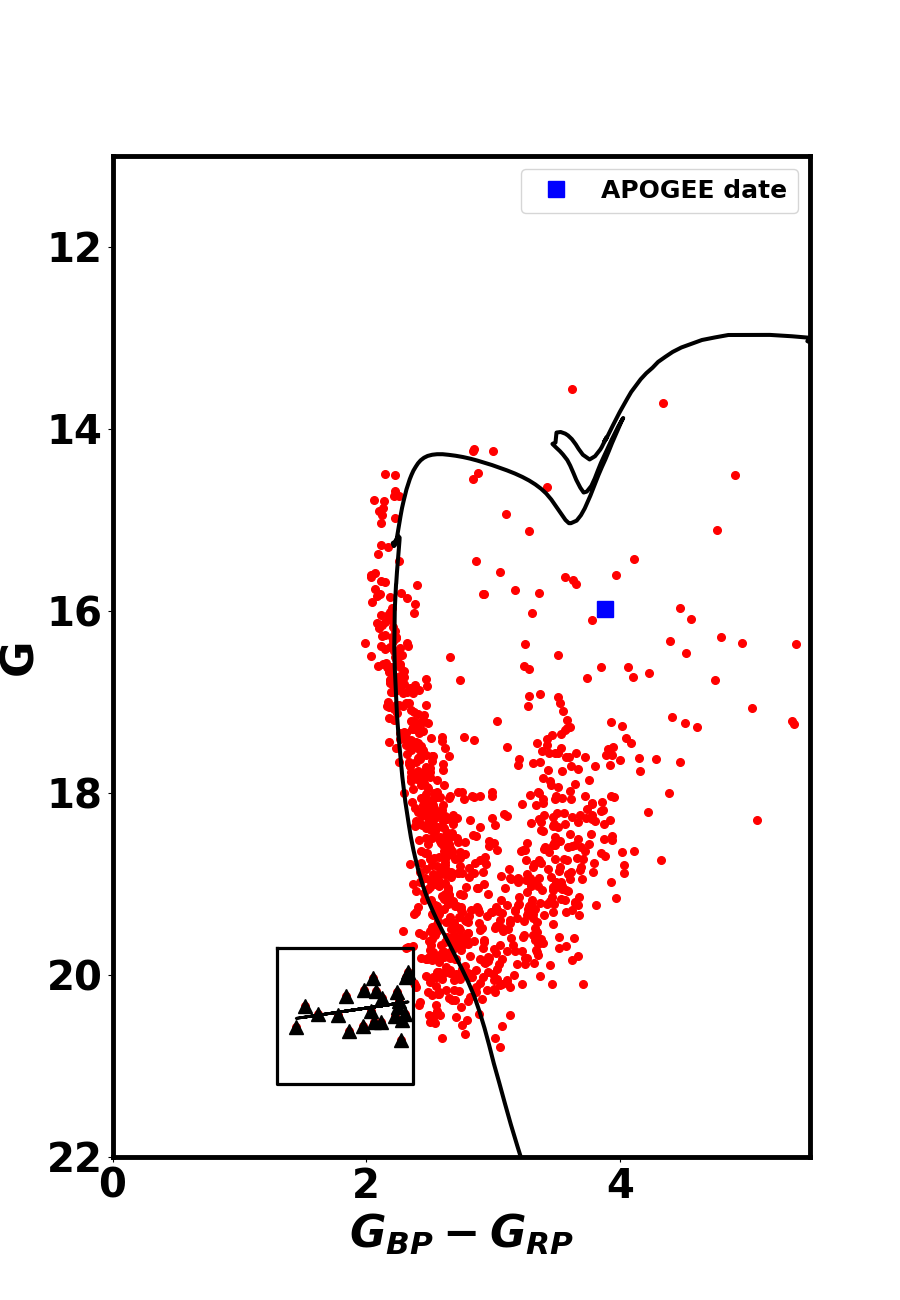}}\hfill
\caption{The CMDs of both clusters with the same isochrone. }
\label{fig:CMD}
\end{figure}

Using the photometric data from Gaia DR3 for stars in Czernik 38, the color-magnitude diagram (CMD) is presented in Fig.~\ref{fig:CMD_new}. The CMD is fitted with theoretical isochrones from \cite{Marigo2017}. \\
We have determined that the intrinsic distance modulus and the color excess, E(G$_{BP}$ - G$_{RP}$), are \dm mag and \CE mag, respectively, see table \ref{tab:phot}. The true distance modulus $\left( m - M \right)_{o}$ and the extinction in the G band ($A_G$) were calculated using the equation provided in Section \ref{sec:extin}. These results correspond to an isochrone-based distance $d_{iso}$ of \distph pc. Fig.~\ref{fig:CMD_new} illustrates the Gaia CMD for Czernik 38, utilizing the broad photometric bands (G, G$_{BP}$, and G$_{RP}$) from the Gaia DR3 data set.
Additionally, the fitted isochrone indicates a cluster age of about \age Myr, with a metallicity of Z=	\Zini~( [M/H]=\MH dex ).
\begin{table*}
\centering
\caption{Gaia Isochrone fit }
\begin{tabular}{| l|c | c | c |} 
 \hline
 The Name      & $E(G_{BP}-G_{RP})$ & $(G-M_G)_o$ & age \\
               & mag.             & mag.        & Myr \\
 \hline
The new Cluster & \CE             & \dm         & \age \\
\hline
Czernik 38 & 2.40 $\pm$0.05 & 12.69$\pm$ 0.08 &  125.0 $\pm$ 12.30 \\
\hline
\end{tabular}
\label{tab:phot}
\end{table*}
\subsection{The Right Branch of CMD} 
The new cluster and Czernik 38 \cite{Nasser2025c} are twin structures and have comparable CMD attributes.
In both clusters, many stars are situated almost parallel to the main sequence on the right side of both Gaia CMDs (right branch stars of CMD, RBCMD); please refer to Fig.~\ref{fig:CMD}.  In our earlier study concerning Czernik 38, we have thoroughly established that these stars are indeed members and are classified as pre-main sequence stars. In particular, we will repeat certain procedures, though not all, to verify the status of these stars. Additionally, we have found the same pattern in our previous research on the King 18 cluster \cite{Nasser2024}, King 13 cluster \cite{Nasser2025b}, and Czernik 38 cluster \cite{Nasser2025c},~ \emph{all of them are young clusters}. Consequently, rather than being a coincidence or  field stars, the existence of this type of star in these clusters must be connected to their youth

\begin{figure*}
\centering
\includegraphics[width=13.5cm, angle=0]{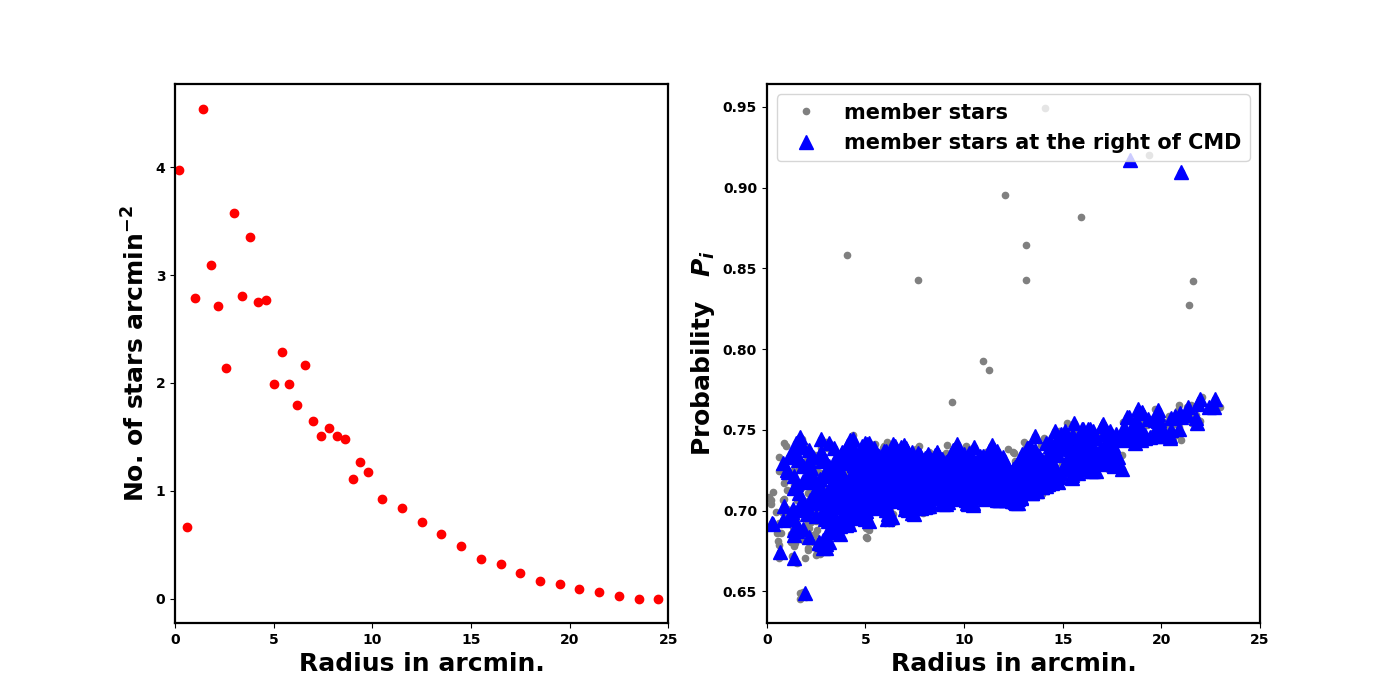}
\caption{The left panel displays the RDP of stars located to the right of CMD.~ Meanwhile, the right panel illustrates the plot of probability in relation to the radius.} 
\label{fig:rcmd}
\end{figure*}

\begin{figure} 
   \centering
   \includegraphics[width=0.95\linewidth, angle=0]{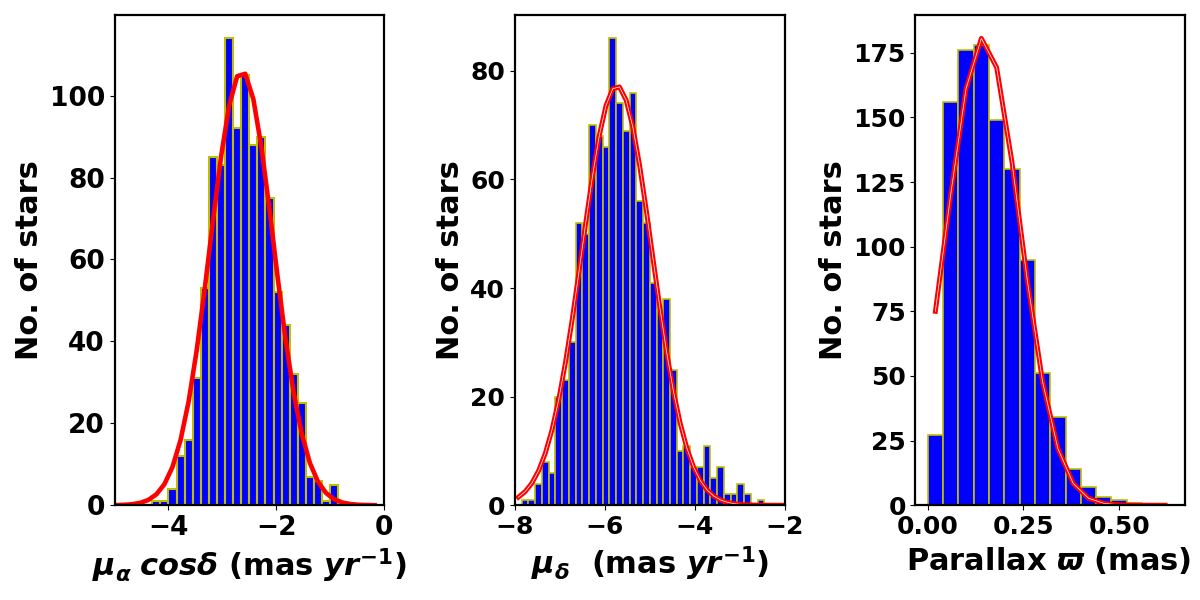}
   \caption{The proper motions and parallaxes histograms of RBCMD stars. Gaussian fits are shown by the solid lines. This figure demonstrates that the  parallaxes and proper motions histograms of RBCMD stars are the same as those of MS stars.} 
   \label{fig:dist_RCMD}
\end{figure}

First, as shown in the left panel of Fig.~\ref{fig:rcmd}, we can get the RDP of these stars. Their overdensity is evident from this diagram.  Moreover, we present the probability in relation to radius as shown in the right panel of Fig  \ref{fig:rcmd}. This figure demonstrates that these stars possess high probability values.  Thus, the two panels of Fig.~\ref{fig:rcmd} confirm that these stars are truly member stars. Moreover, Fig.~\ref{fig:dist_RCMD} demonstrates that the parallaxes and proper motions histograms of RBCMD stars are the same as those of MS stars. 

\begin{figure*}
\centering
\includegraphics[width=8.5cm, angle=0]{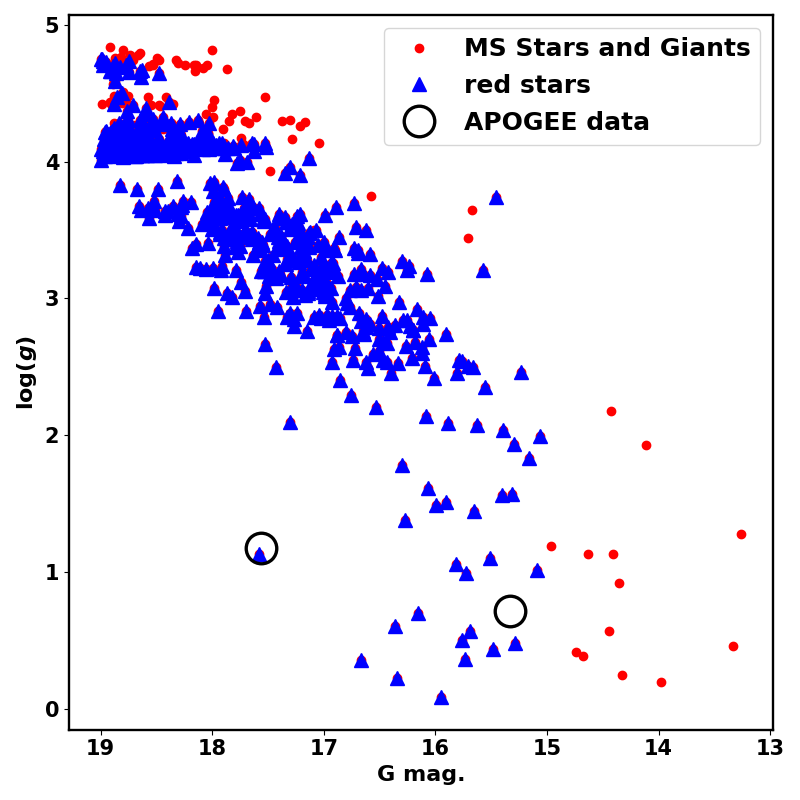}
\caption{The triangles represent the red stars, which are distinguished by their reduced surface gravity and cooler environments compared to main sequence stars. The open circles represent spectral data sourced from \cite{Abdurro2022}, which is consistent with the Gaia DR3 data.
This indicates that these stars are pre-main sequence stars rather than field stars.}
\label{fig:rstars}
\end{figure*}

Although these stars exhibit comparatively lower temperatures, their luminosities are comparable to those of main sequence stars. This implies that they have larger surface areas and, consequently, a greater radius than main sequence stars. This result aligns with the fact that these stars exhibit lower surface gravity than main sequence stars, as illustrated in Fig. \ref{fig:rstars}. The two squares depicted in this figure and Fig. \ref{fig:CMD_new} signify members that possess high-resolution spectral data acquired from \cite{Abdurro2022}, demonstrating a notable correlation with the Gaia DR3 data.

Therefore, these stars might indicate pre-main sequence stars or protostars, with hydrogen burning still not initiated; in other words, they are currently undergoing the formation process.  This suggests that the cluster demonstrates a high rate of star formation, and as a result, it must be a young cluster. 
The prominent feature of star formation in this cluster corresponds with its comparatively youthful age, considerable reddening value, and its unique location within a region of high stellar density and dense gases.\\

In summary, concerning this topic, a thorough investigation into this matter is necessary. To delve deeper and tackle the issue, it is vital to perform an extensive spectroscopic analysis of these stars to gain insights into their characteristics and physical properties. This analysis is fundamental for the research on stellar formation and evolution. \\
\subsection{The Faint Blue Stars}
%
A white dwarf located in a young open cluster represents a stellar remnant that has evolved from a massive progenitor star within a relatively youthful star cluster, marking a rare yet significant discovery for astrophysicists.

As previously stated, the new cluster and Czernik 38 
\cite{Nasser2025c} are twin structures that exhibit similar CMD characteristics.
In current work, we have found that a number of stars appear to be fainter and bluer than the main sequence stars,  which are indicated by stars in the rectangle in Fig.~\ref{fig:CMD}. Moreover, these stars are most probable member stars. These stars could potentially be young white dwarfs based on their locations in the Color-Magnitude Diagram (CMD). 
Interestingly, the correlation between $G$ magnitude and the color $G_{BP} - G_{RP}$ is nearly the same in both clusters, as seen in Fig.~\ref{fig:CMD}, particularly since they share the same age and the same place. 
%
Moreover,  the same pattern is found in the young open clusters Stock 12 and ASCC 113 \cite{Richer2021}.
%
%
\subsection{\textbf{Cluster Mass,  Mass Function and Mass Populations}} \label{sec:mass}
\subsubsection{The cluster Mass}
The masses of individual stars, together with the total mass of the cluster, are vital for grasping the properties of a cluster. Upon completing the isochrone fitting process,  we acquired the absolute magnitude, M$_G$, and the intrinsic color, $G_{BP}-G_{RP}$. Utilizing straightforward polynomial fitting for the calculation of stellar masses may yield inaccurate outcomes because of its inherent limitations. Therefore, we implemented an interpolation routine that incorporates two independent variables.
Specifically, we utilized the \textit{SmoothBivariateSpline} function available in the Python Scipy\footnote{https://scipy.org/} package \cite{SciPy2020} , which enables interpolation with respect to two variables, as the stellar mass is influenced by both the magnitude and the color. 

We utilized $(G_{BP}-G_{RP})_o$ and $M_G$  as independent variables derived from the optimal isochrone to interpolate the masses of individual stars. This methodology allowed us to precisely ascertain the mass of each member within the cluster.
\subsubsection{The Traditional Mass Function}
\begin{figure}
\centering
\subfloat[The Nasser 1 cluster ]{\label{fig:cc}\includegraphics[width=.44\linewidth]{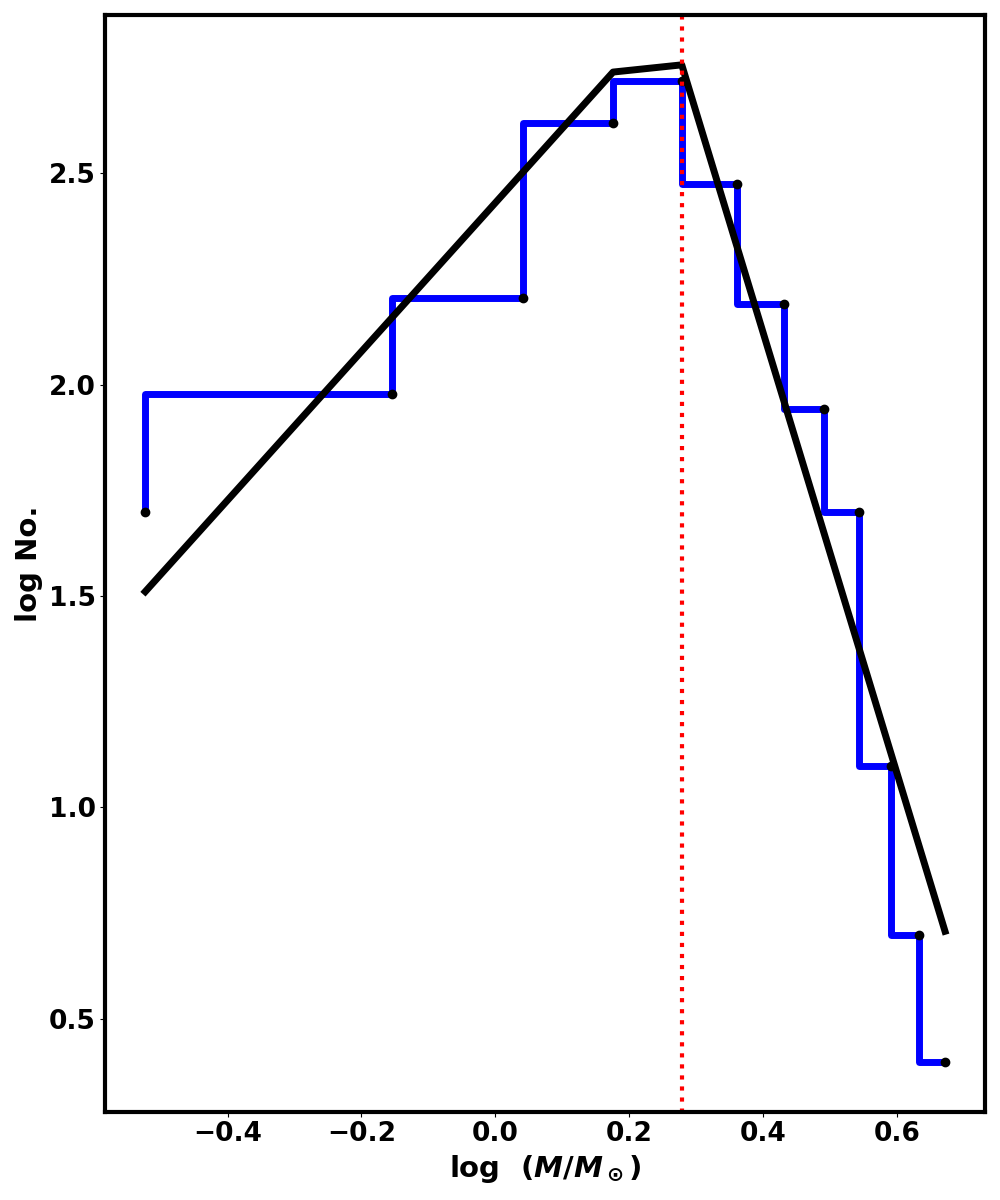}}\hfill
\subfloat[The Czernik 38 cluster]{\includegraphics[width=.44\linewidth]{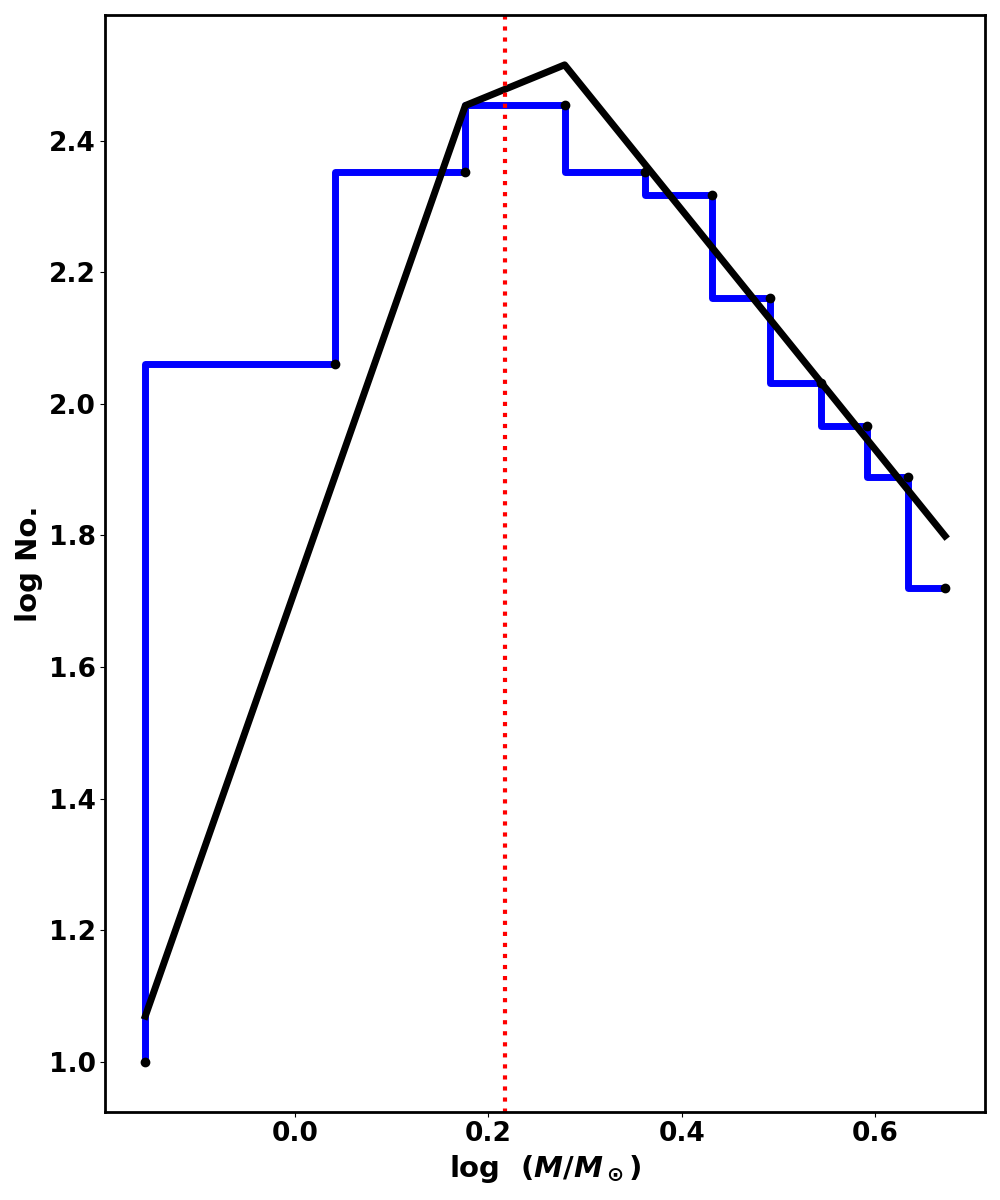}}\hfill
\caption{Salpeter mass function} 
\label{fig:Sal_MF}
\end{figure}

Our study showed that the Salpeter mass function \cite{Salpeter1955}  is inadequate for representing the faint stars of  the Gaia DR3 dataset. Consequently, we chose to adopt a piecewise power law that  incorporates two distinct parts of the power law, as presented in the research by \cite{Almeida2023}. Thus, in our earlier publications \cite{Nasser2025a} and \cite{Nasser2025b}, we depict the mass function (MF) through this piecewise power law, referred to as the modified Salpeter mass function..
%
It can be represented as follows:
%
\begin{equation} 
f(M) = \frac{dN}{dm}  = 
\Bigg \{
\begin{array}{lcc} 
      K_1 \;  m^{-\alpha_1}  &, & \text{if } m \le M_{cr}\\
      K_2 \; m^{-\alpha_2}  &, & \text{if } m >  M_{cr}
\end{array}
\label{eq:mass_fun}    
\end{equation}
under conditions, the function $f(M)$ is continuous :
\begin{equation*}
K_1 \; M_{cr}^{-\alpha_1} = K_2\; M_{cr}^{-\alpha_2}
\end{equation*}
where {dN}/{dm} indicates the number of stars within the mass range $m$ to $m + dm$. The parameters $\alpha_1$ and $\alpha_2$ represent the low and high mass slopes of the mass function, respectively, while $M_{cr}$ marks the critical mass where the slope changes in value and sign. In many open clusters, the hight mass slope $\alpha_2$ is near to Salpeter value of 2.35 \cite{Salpeter1955}.  The fitting of this function is performed using the curve\_fit function of the Scipy python package with $\alpha_1$, $\alpha_2$, $M_{cr}$, $K_1$, and $K_2$ treated as free parameters, constrained by the condition $ f(M^{-}_{cr}) = f(M^{+}_{cr})$, see Table \ref{tab:Sal_MF} and Fig \ref{fig:Sal_MF}.
\subsubsection{The Gaussian Mass Function}
\begin{figure*}
\centering
\begin{tabular}{c c c}
\subfloat[The Gaussian mass function of the \\ ~~Nasser 1 cluster] { \label{fig:Gauss_MF_new} \includegraphics[width=.3\linewidth]{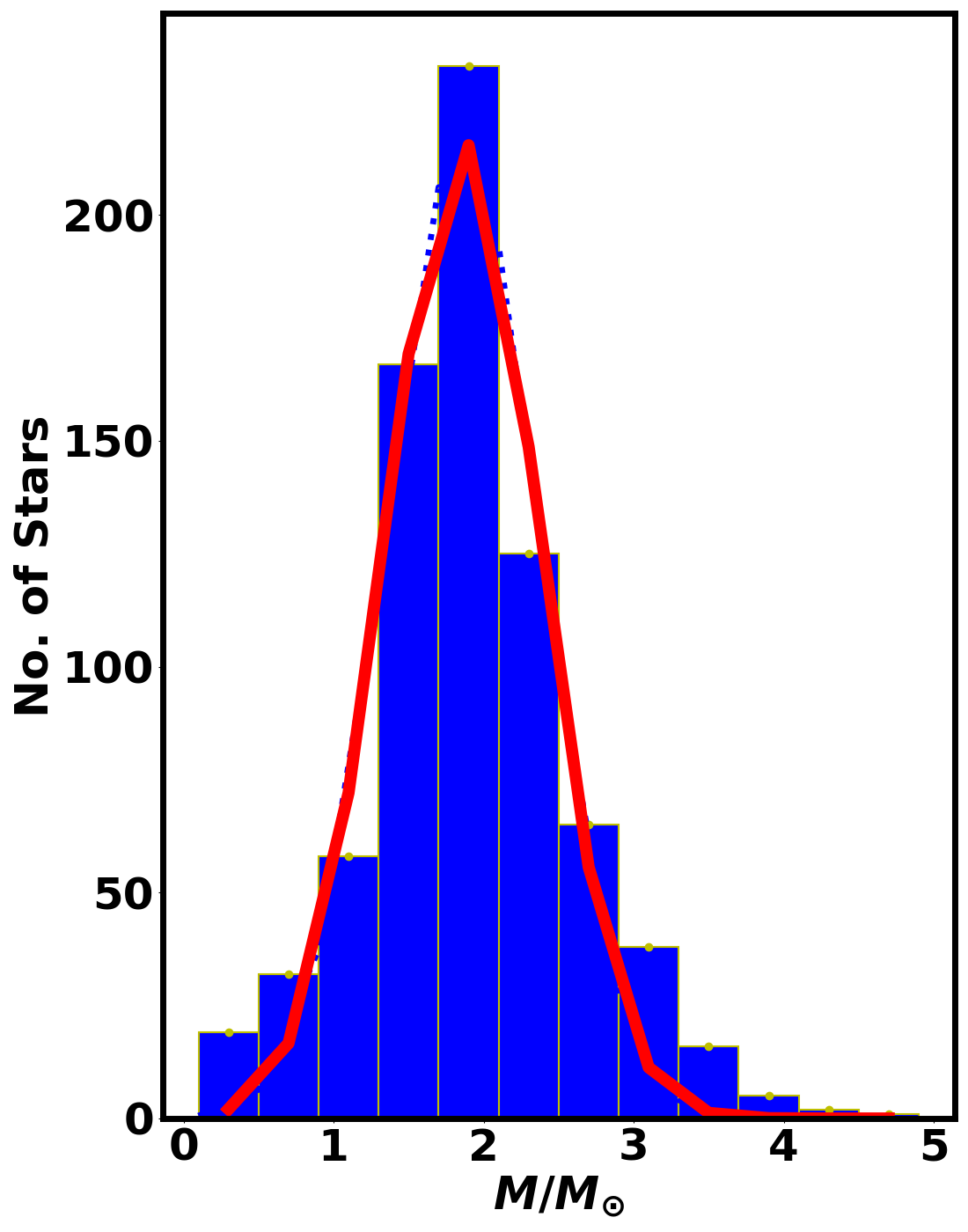} }
\subfloat[The Gaussian mass function of \\~~the Czernik 38 cluster] {\label{fig:Gauss_MF_Cze38} \includegraphics[width=.3\linewidth]{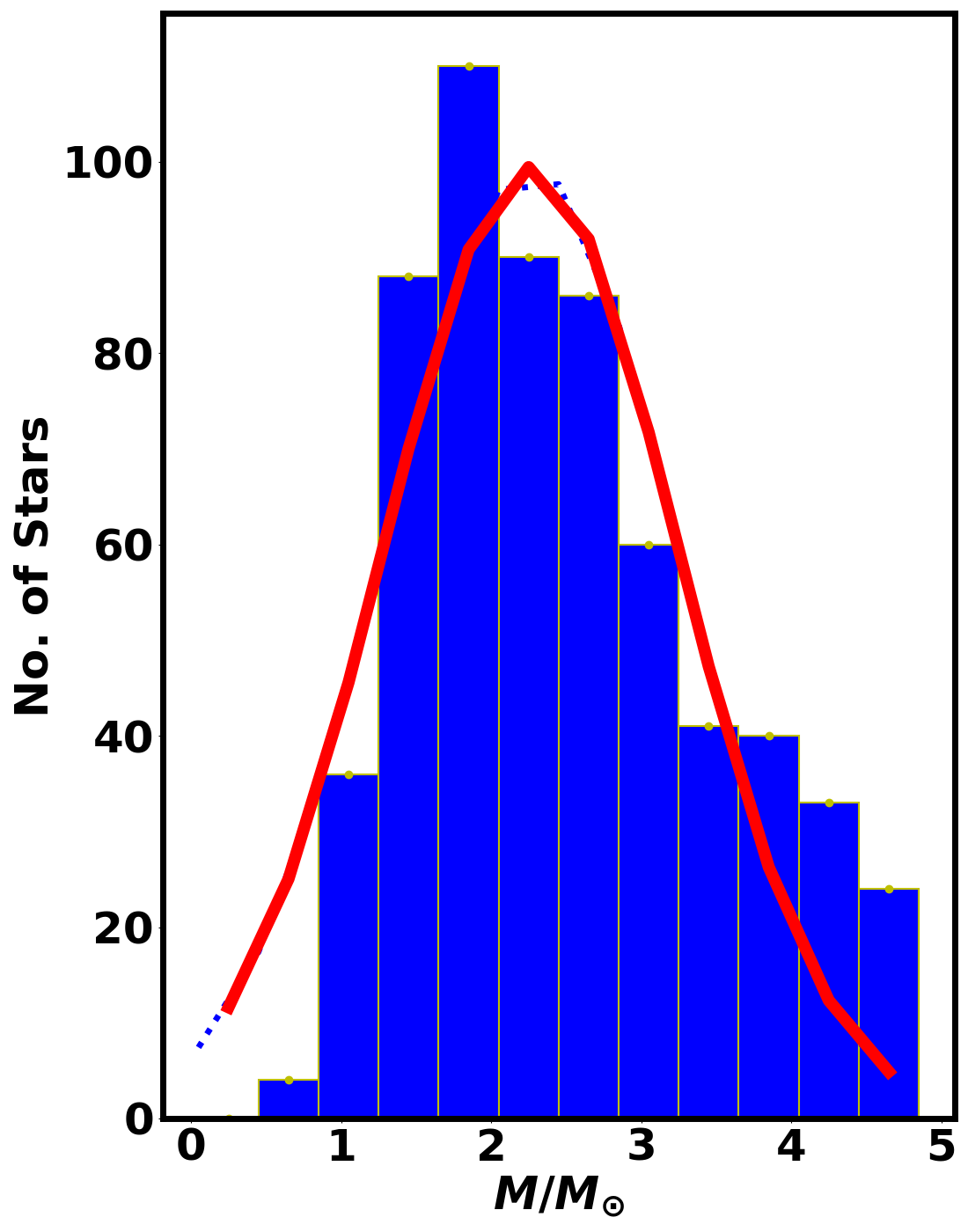} }
\subfloat[The Gaussian mass function for the combined members of the two clusters is displayed in this figure.  \textbf{This mass function fits the data better than two clusters independently.}] { \label{fig:Gauss_MF_all}\includegraphics[width=.3\linewidth]{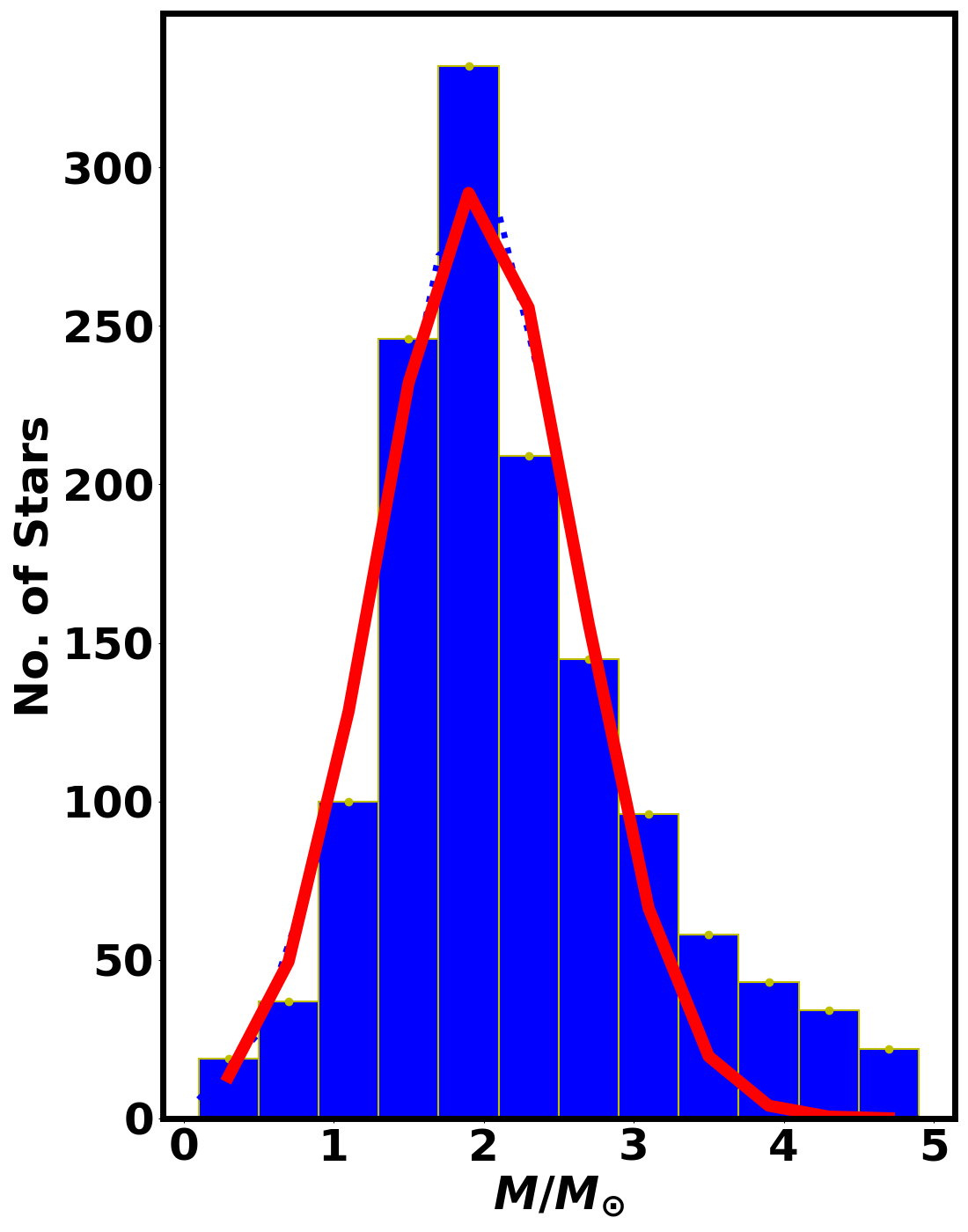} }
\end{tabular}
\caption{The Gaussian mass functions}
\label{fig:Gauss_MF}
\end{figure*}

This research examines the mass distribution by employing an alternative formula for the mass distribution, aiming to identify the most suitable function that accurately represents it and possesses significant physical relevance.  In our earlier work \cite{Nasser2025a,Nasser2025b}, we have introduced the Gaussian distribution, as it represents the optimal fit for the mass function and is more physically relevant.
\begin{equation}
\frac{dN(m)}{dm} =\frac{dN_o}{dm} \times \exp\left({-\dfrac{(m-\mu_m)^2}{2 \; \sigma^2_m} } \right)   
\end{equation}
%
%
where $dN(m)/dm$ is the number of stars having mass m in interval $dm$ and $\mu_m$ is the mean of masses. In this work as well, we perform the same action. Interestingly, however, the mass distribution of this Czernik 38 cluster has Gaussian distributions (Fig.~\ref{fig:Gauss_MF}), see Table \ref{tab:Gauss_MF} for the best fit of the parameters. In other words, the probability of finding a star at mass m in the cluster is represented by the term $dN(m)/dm$. In any case, this equation is more physically meaningful than equation \ref{eq:mass_fun}. However, the theoretical analysis related to this topic is beyond the scope of this paper. 

In this study, we fit the mass distributions of the Czernik 38 cluster and the new cluster using the Gaussian function, as shown in Fig \ref{fig:Gauss_MF}.~\textbf{The two clusters have nearly the same Gaussian fit, see~table~\ref{tab:Gauss_MF}.}
Additionally, to provide further  evidence, we merged the member stars of both clusters and executed a Gaussian fit, with the findings displayed in  \ref{fig:Gauss_MF_all}.  We get a distribution that is more precise than the individual cluster and closely resembles one Gaussian distribution.~ In other words, \textbf{the two clusters have a single Gaussian mass distribution.} 

This section provides a useful practical application of the differences between the Gaussian mass function and the Salpeter mass function \cite{Salpeter1955}, as well as explaining which is more crucial for the star cluster research. 
The slopes of the Salpeter mass function vary between the two clusters, which does not inherently suggest a differing mass distribution. In contrast, the Gaussian function maintains consistency across both clusters, meaning that the two clusters exhibit the same mass distribution, unlike the Salpeter mass function.  This highlights the enhanced significance of the Gaussian function relative to the Salpeter mass function, as it more accurately reflects the mass distribution. The interpretations of these two functions lead to completely distinct results.
\subsection{The Multiple Mass Populations}
\begin{figure}
\centering
\includegraphics[width=8.cm, angle=0]{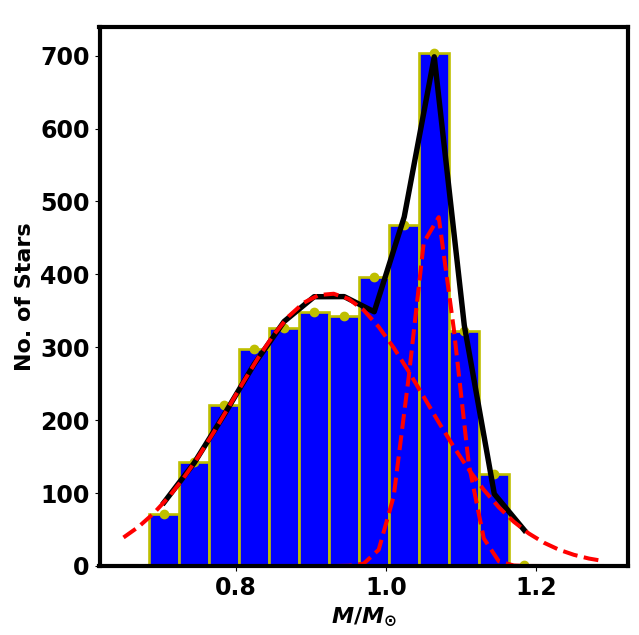}
\caption{The Gaussian mass function  of NGC 6791~\cite{Nasser2025d}.  \textbf{This figure displays the mass distribution as two Gaussian distributions.}  The solid  lines are the fits of two Gaussian mass functions.}
\label{fig:NGC6791_imf}
\end{figure}

Stellar populations provide essential information about the processes involved in star cluster formation and their subsequent evolution. Stellar populations refer to groups of stars that share similar characteristics, primarily age and chemical composition.
In our prior study concerning the thorough examination of the old massive cluster NGC 6791 \cite{Nasser2025d}, Surprisingly, we have found   \emph{two Gaussian mass distributions} for this single cluster; see Fig.~\ref{fig:NGC6791_imf}.
Then, we hereby have presented the concept of multiple mass populations or multiple mass distributions, wherein the stellar populations may not vary in age or chemical composition.  \textbf{Each Gaussian distribution corresponds to a single mass population, which means that the two Gaussian mass distributions are understood as the two mass populations.} 

The massive NGC 6791 cluster consists of two distinct mass populations, whereas this primordial binary cluster only has a single mass population.  The following section will discuss these mass populations in the context of the initial dynamics and formation.
\begin{table*}
\centering
\caption{The two power lows  fit of mass function.}
\begin{tabular}{|c |c|c|c|c|c|}
\hline
The name & $K_1$   & $ K_2 $ & $M_{cr}$   & $\alpha_1$ & $\alpha_2$ \\
  & -   &  -      & $M_{\odot}$ &     -       & - \\
\hline
Czernik 38 &3.4 $\pm$ .09     &  2.35 $\pm$ .04  &  0.373 $\pm$ .02        &-1.49 $\pm$ .03      & 2.1 $\pm$ .07 \\
\hline
The Nasser 1 cluster &3.4 $\pm$ .11     &  3.33 $\pm$ .06  &  0.31 $\pm$ .03        &-1.56 $\pm$ .05      & 1.6 $\pm$ .07 \\
\hline
\end{tabular}
\label{tab:Sal_MF}
\end{table*}
%
%
%
\begin{table}
\centering
\caption{The gaussian fits of mass function.}
\scalebox{0.91}
{\begin{tabular}{|l|c|c|c|}
\hline
The name & $N_o$    & $\mu_m$     & $\sigma_m$    \\
 &stars    & $M_{\odot}$ &  $M_{\odot}$   \\
\hline
Czernik 38& 57.31 $\pm$ 8.53 & 2.11 $\pm$ .06        &0.81 $\pm$ .02     \\
\hline
The Nasser 1 cluster & 111.0 $\pm$ 15.3 & 2.26 $\pm$ .08       &0.71 $\pm$ .03     \\
\hline
\end{tabular}
}
\label{tab:Gauss_MF}
\end{table}
\section{The Morphology of the clusters and The Initial dynamics} \label{sec:dynamics_morphology}
The phenomenon of two-body relaxation \cite{Spitzer1987} leads to mass segregation within star clusters, resulting in the inward movement of massive stars while lighter stars migrate outward. This process facilitates the stripping of low-mass stars into tidal tails due to the gravitational influence of the galaxy \cite{kupper2010tidal}. This is so-called dynamical mass segregation.

However, we possess an alternative perspective contrary to that. Essentially, displacement is dynamically affected by acceleration and the initial parameters of both position and velocity, rather than by  \emph{the mass of the object} as is commonly perceived.  For example, free fall towards the Earth, regardless of air resistance, does not depend on the mass of the object but is instead governed by gravitational acceleration. In other words, objects with different masses will reach the Earth at the same time, provided they have the same initial position and velocity,  as in this equation:
\begin{equation}
 \vec{\boldsymbol{r}(t)} =   \left(\, \vec{\boldsymbol{r_o}} \;+\; \;t\;  \vec{\boldsymbol{v_o}(t)} \,\right) \;+\; \frac{1}{2} \;t^2\;  \vec{\boldsymbol{a}(t)}
\end{equation}
In this context, $\vec{\boldsymbol{r}(t)}$ denotes the current position, $\vec{\boldsymbol{r_o}} $  represents the initial position,  $\vec{\boldsymbol{v_o}(t)}$  signifies the initial velocity vector, and $\vec{\boldsymbol{a}}(t)$ indicates the acceleration vector.

In a comparable way,  the stars located in outer part of  the cluster  experience weaker gravitational binding compared to the other stars. When these  stars are subjected to gravitational forces or differential rotation, they may be influenced by tidal forces, potentially leading to an elongation of the cluster's shape or the formation of a tidal tail, irrespective of their masses.  In other terms, the tidal force affects the stars located in the outer areas of the cluster that are subjected to these force, irrespective of their masses.

In the context of differential rotation, the leading part of a cluster, which experiences reduced gravitational binding, thus moves with a heightened orbital acceleration, regardless of the stars' masses, leading to a tidal effect aligned with the direction of orbital motion.
%
%
\subsection{The morphology of the two clusters and their complex tides}
One of the key characteristics of open clusters (OCs) is the presence of morphological features, like elongations or tidal tails. These tidal tail represents stellar formations that extend from the cluster's main body.  These morphological features primarily indicate the existence of the tidal effect, which may arise from factors such as the tidal influence of a massive object, disk shock, or differential rotation.  Furthermore, the orientation of the elongated structure or tail may reflect the direction of the tide.  Consequently, it is imperative to recognize the direction of the tide, as it is essential to understand the nature of the tidal force. For instance, if the elongation or tail aligns with the direction of orbital motion, then the cluster experiences the effects of differential rotation.

In our case study, this binary cluster system demonstrates a distinctive complexity and impressive tidal phenomena. There are two forms of tides effects \textbf{:} \textbf{(i)}~ the first one is tidal due to  the differential rotation, which aligns with the system's motion, and \textbf{(ii)}~ the second is the tide caused by a massive object, such as a giant cloud or one of the spiral arms. This tide results in an elongated structure oriented toward that massive object.  

\begin{figure}
\centering
\subfloat[The point vector diagram (PVD) of the Nasser 1 cluster illustrates an elongation in the direction of motion due to the differential rotation, as well as an elongation towards a massive object, see Fig. \ref{fig:surf_map} ]{\label{fig:PMV_new}\includegraphics[width=.44\linewidth]{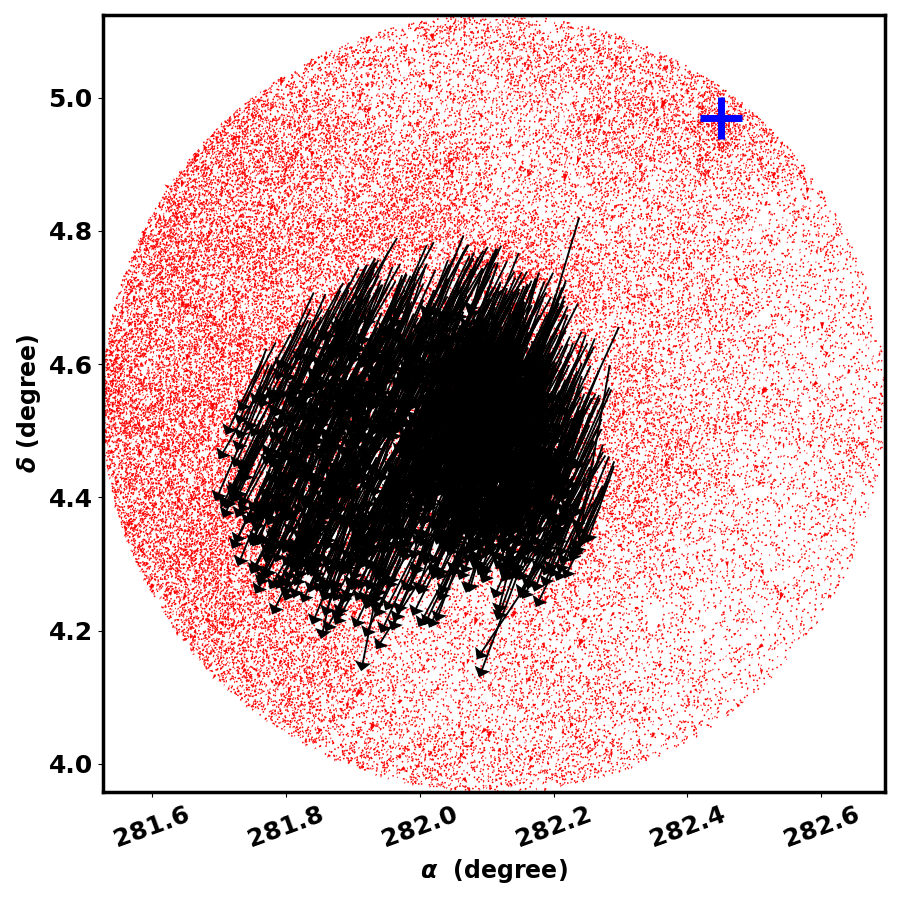}}\hfill
\subfloat[The point vector diagram (PVD) of the Czernik 38 cluster illustrates elongation towards orbital motion as a result of differential rotation.]{\label{fig:PMV_cz38}\includegraphics[width=.44\linewidth]{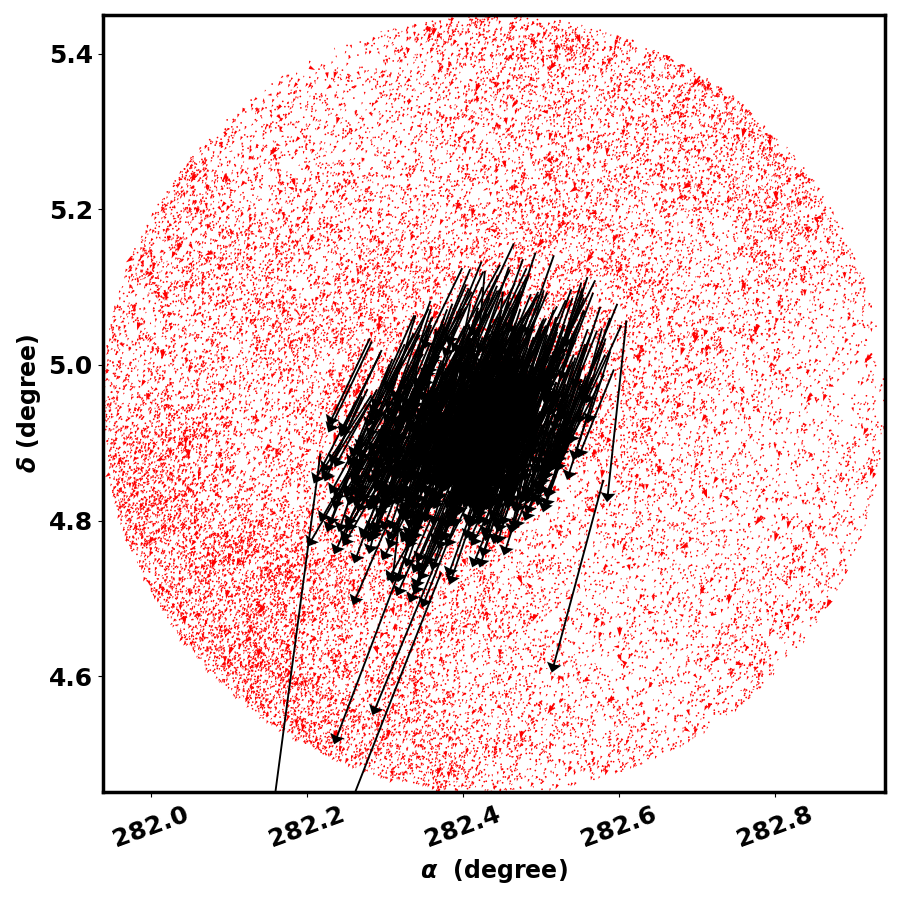}}\hfill
\caption{The point vector diagram (PVD) of both clusters}
\label{fig:PMV}
\end{figure}

The first form, as our previous study revealed a significant tidal tail located outside the radius of the cluster, oriented in the direction of motion as in the case of the open King 13 cluster \cite{Nasser2025b}. Moreover, we have also found a leading tail in Czernik 38 cluster \cite{Nasser2025c}, which is  aligned with the direction of orbital motion , see Fig.~\ref{fig:PMV_cz38}.  Moreover, the Nasser 1 cluster is also influenced by the differential rotation, as noted by Czernik 38, and in the same direction, which offers further evidence that they form a binary cluster.

 This newly identified cluster exhibits this second form of tide; refer to Fig.~\ref{fig:PMV_new}, which is influenced by its proximity to the Carina-Sagittarius arm, with the elongation oriented in the direction of this tidal force. This implies that this star cluster experiences two intriguing and distinct forms of of tidal phenomena.
%
\subsection{The Initial Dynamics}
The morphology of clusters and their Gaussian mass functions or mass populations are essential in analyzing the initial dynamics of any cluster system, its dynamical evolution, and the related dynamics of galaxies.
%

This impressive outcome reveals that, even though there are two clusters, they exhibit a singular mass distribution or mass population. This suggests they were formerly a single cluster that has been violently torn apart by tidal forces, as  represented by the differential  rotation  during its orbital journey.  This is evident from their elongated structures aligned with the direction of orbital motion. 

In comparison, the NGC 6791 cluster is a single cluster with two mass populations. This suggests that they were initially two clusters that were brought together by the gravitational potential and are located far from any disruptive forces.

As previously mentioned, the Gaussian mass function or mass populations serves as a crucial tool for investigating the initial dynamics of any binary or any system of clusters, or even a single cluster, and is considered to be more realistic and accurate than the Salpeter function. The primordial binaries star cluster require a reassessment through the application of a Gaussian function to ascertain if they were previously a singular system that was turn apart  by gravitational tides, or if they were binary from their inception. This approach enables us to investigate not only the dynamics of these clusters but also the dynamics of the Galaxy.
%
%
\newpage
\section{Summary and Conclusions} 
\label{sec:summary}
In our earlier accepted  work regarding the comprehensive analysis of the Czernik 38 cluster through Gaia DR3 data, we discovered a clump of stars located at $
\alpha=$\ra and $\delta=$\dec approximately 32 arcmin from its center. We proposed that this clump may represent a new open stars cluster and suspect it could be a companion to Czernik 38, as a binary open cluster. \\
Consequently, in this ongoing investigation, we aim to examine this issue in detail and determine the physical parameters of this Nasser 1 cluster using Gaia data, first to determine if it exhibits a King profile and CMD. Subsequently, we will compare its parameters with those of Czernik 38 to assess binarity. The main findings are the following.
\begin{enumerate}[ itemsep=.8em]
%
\item [\textbf{1)}] The RDP of the Nasser 1 cluster fits appropriately with the King model. This is the initial step to ensure that it is indeed a open cluster or not.  We have identified \Ns member stars ($N_{cl}$) within a cluster radius of \rt arcmin. 

\item [\textbf{2)}]   The newly identified cluster exhibits  a well-defined CMD. Our analysis revealed that the distance modulus of the cluster was \dm, corresponding to \distph pc. Additionally, we determined that the color excess $E(G_{BP}~-~G_{RP})$ is \CE. \textbf{This cluster exhibits a significant reddening value similar to that of Czernik 38.} 
\item [\textbf{3)}]  The Nasser 1 cluster alongside Czernik 38 represents twin structures that possess the same CMD attributes. On the right side of both the Gaia CMDs, numerous high probability stars are located almost parallel to the main sequence (right branch stars of CMD, RBCMD). These stars may represent a certain category of pre-main sequence stars, particularly as they are predominantly observed in young clusters. To address this issue and to explore this matter in more  depth, it is crucial and essential to conduct a comprehensive spectroscopic analysis of these stars. \\
Furthermore, in both  CMDs  of the two clusters,  we have identified that a number of stars appear fainter and bluer compared to the main sequence stars.   Additionally,  these stars  are most probable member stars.  Their positions in the CMDs, together with their considerable probability values in both clusters,  suggest that they might be young white dwarfs.

\item [\textbf{4)}]  The components of proper motion ($\mu_{\alpha} \cos \delta$, $\mu_{\delta}$) and the parallax ($\varpi$) were evaluated as \pma\ mas yr$^{-1}$, \pmd\ mas yr$^{-1}$ (~the tangential velocity is \Vt~km/s~), and \parx mas, respectively.

The mean distance inferred from the parallax measurements is nearly \dist pc, which is in agreement with the photometric data from Gaia, within the errors.  Moreover,  we have identified \Nrv~  member stars with radial velocity data with average value \Vr km/s, comparable to the Czernik 38 value of 46.1 $\pm$ 47.86. 
\item [\textbf{5)}]Our primary conclusion, derived from the analysis of the parameters of the new star cluster in comparison to the Czernik 38 cluster, indicates that they share the same age, distance, and proper motion. They also share similar reddening levels and both exhibit a high rate of star formation. Thus, they together create a  fascinating primordial binary cluster system located  at the outskirts of the Carina-Sagittarius arm.\\
\item [\textbf{6)}] 
The Gaussian mass function is considered to be more physically realistic than the Salpeter mass function, as it more precisely represents the distribution of masses, yielding useful insights in the studies of open star clusters. One outcome of employing the Gaussian distribution is the concept of mass population. In addition, the mass distribution or the mass population can yield crucial information regarding the early dynamics of cluster formation.\\
Even though this system has two clusters, it only has  a single Gaussian mass distribution  or an one mass population, indicating that it was once a single cluster that has been violently torn apart by strong tidal forces during its orbital journey. 
\item [\textbf{7)}]
This binary cluster system exhibits a unique and astonishing complex tides. There are two forms of tides.  In the first form, we have found a  elongated structure   in Czernik 38 cluster, which is  aligned with the direction of orbital motion, \textbf{This type of tide is due to the differential rotation.}

This newly identified cluster exhibits this second form of tide, which is influenced by its proximity to the Carina-Sagittarius arm, with the elongation oriented in the direction of this tidal force.  Furthermore, this cluster is also affected by the  tide as Czernik 38 due to the differential rotation and in the same direction, which provides additional evidence that they constitute a binary cluster.
\end{enumerate}
\section*{Data Availability} 
\begin{itemize}
    \item \textbf{Gaia DR3  data~~~~~~~~~~~~~~:} are available for free in webpage \textcolor{blue}{\textbf{https://vizier.cds.unistra.fr/ }}
\end{itemize}  


%
%
%
\section*{Acknowledgements}
This work has made use of data from the European Space Agency (ESA) space mission Gaia. Gaia data are being processed by the Gaia Data Processing and Analysis Consortium (DPAC). Funding for the DPAC is provided by national institutions, in particular the institutions participating in the Gaia MultiLateral Agreement (MLA). The Gaia mission website is \textcolor{blue}{https://www.cosmos.esa.int/gaia}. The Gaia archive website is \textcolor{blue}{https://archives.esac.esa.int/gaia}. 

The authors are pretty thankful to Python community groups for large efforts especially for Matplotlib, Numpy, Scipy and Astropy etc. Their efforts have contributed to making data analysis easier as well as representing it graphically in a creative way.
%

%
\section*{Declaration Section }
\subsection*{Author Contributions}
Nasser M. Ahmed~~:~~~100\%
%
\subsection*{Funding}
Open access funding provided by The Science, Technology \& Innovation Funding Authority (STDF) in cooperation with The Egyptian Knowledge Bank (EKB)
\subsection*{conflict of interest statement :} Not applicable
\bibliography{sample} 

\begin{thebibliography}{10}
\urlstyle{rm}
\expandafter\ifx\csname url\endcsname\relax
  \def\url#1{\texttt{#1}}\fi
\expandafter\ifx\csname urlprefix\endcsname\relax\def\urlprefix{URL }\fi
\expandafter\ifx\csname doiprefix\endcsname\relax\def\doiprefix{DOI: }\fi
\providecommand{\bibinfo}[2]{#2}
\providecommand{\eprint}[2][]{\url{#2}}

\bibitem{Madore2022}
\bibinfo{author}{{Madore}, B.~F.}, \bibinfo{author}{{Freedman}, W.~L.}, \bibinfo{author}{{Lee}, A.~J.} \& \bibinfo{author}{{Owens}, K.}
\newblock \bibinfo{journal}{\bibinfo{title}{{Milky Way Zero-point Calibration of the JAGB Method: Using Thermally Pulsing AGB Stars in Galactic Open Clusters}}}.
\newblock {\emph{\JournalTitle{\apj}}} \textbf{\bibinfo{volume}{938}}, \bibinfo{pages}{125}, \doiprefix\url{10.3847/1538-4357/ac92fd} (\bibinfo{year}{2022}).
\newblock \eprint{2209.08127}.

\bibitem{Carraro2007}
\bibinfo{author}{{Carraro}, G.}, \bibinfo{author}{{Geisler}, D.}, \bibinfo{author}{{Villanova}, S.}, \bibinfo{author}{{Frinchaboy}, P.~M.} \& \bibinfo{author}{{Majewski}, S.~R.}
\newblock \bibinfo{journal}{\bibinfo{title}{{Old open clusters in the outer Galactic disk}}}.
\newblock {\emph{\JournalTitle{\aap}}} \textbf{\bibinfo{volume}{476}}, \bibinfo{pages}{217--227}, \doiprefix\url{10.1051/0004-6361:20078113} (\bibinfo{year}{2007}).
\newblock \eprint{0709.2126}.

\bibitem{Cantat-Gaudin2020}
\bibinfo{author}{{Cantat-Gaudin}, T.} \& \bibinfo{author}{{Anders}, F.}
\newblock \bibinfo{journal}{\bibinfo{title}{{Clusters and mirages: cataloguing stellar aggregates in the Milky Way}}}.
\newblock {\emph{\JournalTitle{\aap}}} \textbf{\bibinfo{volume}{633}}, \bibinfo{pages}{A99}, \doiprefix\url{10.1051/0004-6361/201936691} (\bibinfo{year}{2020}).
\newblock \eprint{1911.07075}.

\bibitem{Lada2003}
\bibinfo{author}{{Lada}, C.~J.} \& \bibinfo{author}{{Lada}, E.~A.}
\newblock \bibinfo{journal}{\bibinfo{title}{{Embedded Clusters in Molecular Clouds}}}.
\newblock {\emph{\JournalTitle{\araa}}} \textbf{\bibinfo{volume}{41}}, \bibinfo{pages}{57--115}, \doiprefix\url{10.1146/annurev.astro.41.011802.094844} (\bibinfo{year}{2003}).
\newblock \eprint{astro-ph/0301540}.

\bibitem{Nasser2025c}
\bibinfo{author}{{Ahmed}, N.~M.}
\newblock \bibinfo{journal}{\bibinfo{title}{{A detailed analysis of the Czernik 38 cluster and its associated tidal tail, utilizing Gaia DR3 and 2MASS }}}.
\newblock {\emph{\JournalTitle{Scientific Reports; acceptted, Submission ID: "8a531eac-05f0-409a-8df0-c98ecc789b2d"}}} \textbf{\bibinfo{volume}{-}}, \doiprefix\url{https://doi.org/10.21203/rs.3.rs-7600993/v1} (\bibinfo{year}{2025}).

\bibitem{Darma2021}
\bibinfo{author}{{Darma}, R.}, \bibinfo{author}{{Arifyanto}, M.~I.} \& \bibinfo{author}{{Kouwenhoven}, M.~B.~N.}
\newblock \bibinfo{journal}{\bibinfo{title}{{The formation of binary star clusters in the Milky Way and Large Magellanic Cloud}}}.
\newblock {\emph{\JournalTitle{\mnras}}} \textbf{\bibinfo{volume}{506}}, \bibinfo{pages}{4603--4620}, \doiprefix\url{10.1093/mnras/stab1931} (\bibinfo{year}{2021}).
\newblock \eprint{2107.03013}.

\bibitem{Arnold2017}
\bibinfo{author}{{Arnold}, B.}, \bibinfo{author}{{Goodwin}, S.~P.}, \bibinfo{author}{{Griffiths}, D.~W.} \& \bibinfo{author}{{Parker}, R.~J.}
\newblock \bibinfo{journal}{\bibinfo{title}{{How do binary clusters form?}}}
\newblock {\emph{\JournalTitle{\mnras}}} \textbf{\bibinfo{volume}{471}}, \bibinfo{pages}{2498--2507}, \doiprefix\url{10.1093/mnras/stx1719} (\bibinfo{year}{2017}).
\newblock \eprint{1707.02990}.

\bibitem{Mora2019}
\bibinfo{author}{{Mora}, M.~D.}, \bibinfo{author}{{Puzia}, T.~H.} \& \bibinfo{author}{{Chanam{\'e}}, J.}
\newblock \bibinfo{journal}{\bibinfo{title}{{On collision course: The nature of the binary star cluster NGC2006/SL 538}}}.
\newblock {\emph{\JournalTitle{\aap}}} \textbf{\bibinfo{volume}{622}}, \bibinfo{pages}{A65}, \doiprefix\url{10.1051/0004-6361/201834103} (\bibinfo{year}{2019}).

\bibitem{Dieball2002}
\bibinfo{author}{{Dieball}, A.}, \bibinfo{author}{{M{\"u}ller}, H.} \& \bibinfo{author}{{Grebel}, E.~K.}
\newblock \bibinfo{journal}{\bibinfo{title}{{A statistical study of binary and multiple clusters in the LMC}}}.
\newblock {\emph{\JournalTitle{\aap}}} \textbf{\bibinfo{volume}{391}}, \bibinfo{pages}{547--564}, \doiprefix\url{10.1051/0004-6361:20020815} (\bibinfo{year}{2002}).
\newblock \eprint{astro-ph/0206364}.

\bibitem{Bergh1996}
\bibinfo{author}{{van den Bergh}, S.}
\newblock \bibinfo{journal}{\bibinfo{title}{{Mergers of Globular Clusters}}}.
\newblock {\emph{\JournalTitle{\apjl}}} \textbf{\bibinfo{volume}{471}}, \bibinfo{pages}{L31}, \doiprefix\url{10.1086/310331} (\bibinfo{year}{1996}).
\newblock \eprint{astro-ph/9609095}.

\bibitem{Fuente2009}
\bibinfo{author}{{de La Fuente Marcos}, R.} \& \bibinfo{author}{{de La Fuente Marcos}, C.}
\newblock \bibinfo{journal}{\bibinfo{title}{{Double or binary: on the multiplicity of open star clusters}}}.
\newblock {\emph{\JournalTitle{\aap}}} \textbf{\bibinfo{volume}{500}}, \bibinfo{pages}{L13--L16}, \doiprefix\url{10.1051/0004-6361/200912297} (\bibinfo{year}{2009}).
\newblock \eprint{0904.4017}.

\bibitem{Leon1999}
\bibinfo{author}{{Leon}, S.}, \bibinfo{author}{{Bergond}, G.} \& \bibinfo{author}{{Vallenari}, A.}
\newblock \bibinfo{journal}{\bibinfo{title}{{Interacting star clusters in the Large Magellanic Cloud. Overmerging problem solved by cluster group formation}}}.
\newblock {\emph{\JournalTitle{\aap}}} \textbf{\bibinfo{volume}{344}}, \bibinfo{pages}{450--458}, \doiprefix\url{10.48550/arXiv.astro-ph/9812112} (\bibinfo{year}{1999}).
\newblock \eprint{astro-ph/9812112}.

\bibitem{Rozhavskii1976}
\bibinfo{author}{{Rozhavskii}, F.~G.}, \bibinfo{author}{{Kuz'mina}, V.~A.} \& \bibinfo{author}{{Vasilevskii}, A.~E.}
\newblock \bibinfo{journal}{\bibinfo{title}{{Statistical approach toward determining the multiplicity of open stellar clusters}}}.
\newblock {\emph{\JournalTitle{Astrophysics}}} \textbf{\bibinfo{volume}{12}}, \bibinfo{pages}{204--209}, \doiprefix\url{10.1007/BF01002037} (\bibinfo{year}{1976}).

\bibitem{Bhatia1988}
\bibinfo{author}{{Bhatia}, R.~K.} \& \bibinfo{author}{{Hatzidimitriou}, D.}
\newblock \bibinfo{journal}{\bibinfo{title}{{Binary star clusters in the Large Magellanic Cloud.}}}
\newblock {\emph{\JournalTitle{\mnras}}} \textbf{\bibinfo{volume}{230}}, \bibinfo{pages}{215--221}, \doiprefix\url{10.1093/mnras/230.2215} (\bibinfo{year}{1988}).

\bibitem{Pavlovskaya1989}
\bibinfo{author}{{Pavlovskaya}, E.~D.} \& \bibinfo{author}{{Filippova}, A.~A.}
\newblock \bibinfo{journal}{\bibinfo{title}{{Groups of Stars with Common Motion in the Galaxy - Groups of Stars in Luminosity Classes i and II - Comparison with Groups of Longperiod Cepheids and Open Clusters}}}.
\newblock {\emph{\JournalTitle{\sovast}}} \textbf{\bibinfo{volume}{33}}, \bibinfo{pages}{602} (\bibinfo{year}{1989}).

\bibitem{Hatzidimitriou1990}
\bibinfo{author}{{Hatzidimitriou}, D.} \& \bibinfo{author}{{Bhatia}, R.~K.}
\newblock \bibinfo{journal}{\bibinfo{title}{{Cluster pairs in the Small Magellanic Cloud.}}}
\newblock {\emph{\JournalTitle{\aap}}} \textbf{\bibinfo{volume}{230}}, \bibinfo{pages}{11--15} (\bibinfo{year}{1990}).

\bibitem{Loktin1997}
\bibinfo{author}{{Loktin}, A.~V.}
\newblock \bibinfo{journal}{\bibinfo{title}{{The Selection of Probable Multiple Open Clusters in Our Galaxy}}}.
\newblock {\emph{\JournalTitle{Astronomical and Astrophysical Transactions}}} \textbf{\bibinfo{volume}{14}}, \bibinfo{pages}{181--193}, \doiprefix\url{10.1080/10556799708202989} (\bibinfo{year}{1997}).

\bibitem{Subramaniam1995}
\bibinfo{author}{{Subramaniam}, A.}, \bibinfo{author}{{Gorti}, U.}, \bibinfo{author}{{Sagar}, R.} \& \bibinfo{author}{{Bhatt}, H.~C.}
\newblock \bibinfo{journal}{\bibinfo{title}{{Probable binary open star clusters in the Galaxy.}}}
\newblock {\emph{\JournalTitle{\aap}}} \textbf{\bibinfo{volume}{302}}, \bibinfo{pages}{86} (\bibinfo{year}{1995}).

\bibitem{Dieball1998a}
\bibinfo{author}{{Dieball}, A.} \& \bibinfo{author}{{Grebel}, E.~K.}
\newblock \bibinfo{title}{{Binary clusters in the Magellanic Clouds I: The Large Magellanic Cloud}}.
\newblock In \emph{\bibinfo{booktitle}{Astronomische Gesellschaft Abstract Series}}, vol.~\bibinfo{volume}{14} of \emph{\bibinfo{series}{Astronomische Gesellschaft Abstract Series}}, \bibinfo{pages}{134--134} (\bibinfo{year}{1998}).

\bibitem{Dieball1998b}
\bibinfo{author}{{Dieball}, A.} \& \bibinfo{author}{{Grebel}, E.~K.}
\newblock \bibinfo{title}{{Binary clusters in the Magellanic Clouds II: The Small Magellanic Cloud}}.
\newblock In \emph{\bibinfo{booktitle}{Astronomische Gesellschaft Abstract Series}}, vol.~\bibinfo{volume}{14} of \emph{\bibinfo{series}{Astronomische Gesellschaft Abstract Series}}, \bibinfo{pages}{135--135} (\bibinfo{year}{1998}).

\bibitem{Bekki2004}
\bibinfo{author}{{Bekki}, K.}, \bibinfo{author}{{Beasley}, M.~A.}, \bibinfo{author}{{Forbes}, D.~A.} \& \bibinfo{author}{{Couch}, W.~J.}
\newblock \bibinfo{journal}{\bibinfo{title}{{Formation of Star Clusters in the Large Magellanic Cloud and Small Magellanic Cloud. I. Preliminary Results on Cluster Formation from Colliding Gas Clouds}}}.
\newblock {\emph{\JournalTitle{\apj}}} \textbf{\bibinfo{volume}{602}}, \bibinfo{pages}{730--737}, \doiprefix\url{10.1086/381171} (\bibinfo{year}{2004}).
\newblock \eprint{astro-ph/0312112}.

\bibitem{GaiaCollaboration2023}
\bibinfo{author}{{Gaia Collaboration}} \emph{et~al.}
\newblock \bibinfo{journal}{\bibinfo{title}{{Gaia Data Release 3. Summary of the content and survey properties}}}.
\newblock {\emph{\JournalTitle{\aap}}} \textbf{\bibinfo{volume}{674}}, \bibinfo{pages}{A1}, \doiprefix\url{10.1051/0004-6361/202243940} (\bibinfo{year}{2023}).
\newblock \eprint{2208.00211}.

\bibitem{Pedregosa2011}
\bibinfo{author}{{Pedregosa}, F.} \emph{et~al.}
\newblock \bibinfo{journal}{\bibinfo{title}{{Scikit-learn: Machine Learning in Python}}}.
\newblock {\emph{\JournalTitle{Journal of Machine Learning Research}}} \textbf{\bibinfo{volume}{12}}, \bibinfo{pages}{2825--2830}, \doiprefix\url{10.48550/arXiv.1201.0490} (\bibinfo{year}{2011}).
\newblock \eprint{1201.0490}.

\bibitem{Pera2021}
\bibinfo{author}{{Pera2021}, M.~S.}, \bibinfo{author}{{Perren}, G.~I.}, \bibinfo{author}{{Moitinho}, A.}, \bibinfo{author}{{Navone}, H.~D.} \& \bibinfo{author}{{Vazquez}, R.~A.}
\newblock \bibinfo{journal}{\bibinfo{title}{{pyUPMASK: an improved unsupervised clustering algorithm}}}.
\newblock {\emph{\JournalTitle{\aap}}} \textbf{\bibinfo{volume}{650}}, \bibinfo{pages}{A109}, \doiprefix\url{10.1051/0004-6361/202040252} (\bibinfo{year}{2021}).
\newblock \eprint{2101.01660}.

\bibitem{Song2022}
\bibinfo{author}{{Song}, F.}, \bibinfo{author}{{Esamdin}, A.}, \bibinfo{author}{{Hu}, Q.} \& \bibinfo{author}{{Zhang}, M.}
\newblock \bibinfo{journal}{\bibinfo{title}{{Binary open clusters in the Gaia data}}}.
\newblock {\emph{\JournalTitle{\aap}}} \textbf{\bibinfo{volume}{666}}, \bibinfo{pages}{A75}, \doiprefix\url{10.1051/0004-6361/202243524} (\bibinfo{year}{2022}).
\newblock \eprint{2208.12935}.

\bibitem{Conrad2017}
\bibinfo{author}{{Conrad}, C.} \emph{et~al.}
\newblock \bibinfo{journal}{\bibinfo{title}{{A RAVE investigation on Galactic open clusters . II. Open cluster pairs, groups and complexes}}}.
\newblock {\emph{\JournalTitle{\aap}}} \textbf{\bibinfo{volume}{600}}, \bibinfo{pages}{A106}, \doiprefix\url{10.1051/0004-6361/201630012} (\bibinfo{year}{2017}).

\bibitem{Soubiran2019}
\bibinfo{author}{{Soubiran}, C.} \emph{et~al.}
\newblock \bibinfo{journal}{\bibinfo{title}{{Open cluster kinematics with Gaia DR2 (Corrigendum)}}}.
\newblock {\emph{\JournalTitle{\aap}}} \textbf{\bibinfo{volume}{623}}, \bibinfo{pages}{C2}, \doiprefix\url{10.1051/0004-6361/201834020e} (\bibinfo{year}{2019}).

\bibitem{Liu2019}
\bibinfo{author}{{Liu}, L.} \& \bibinfo{author}{{Pang}, X.}
\newblock \bibinfo{journal}{\bibinfo{title}{{A Catalog of Newly Identified Star Clusters in Gaia DR2}}}.
\newblock {\emph{\JournalTitle{\apjs}}} \textbf{\bibinfo{volume}{245}}, \bibinfo{pages}{32}, \doiprefix\url{10.3847/1538-4365/ab530a} (\bibinfo{year}{2019}).
\newblock \eprint{1910.12600}.

\bibitem{Palma2025}
\bibinfo{author}{{Palma}, T.}, \bibinfo{author}{{Coenda}, V.}, \bibinfo{author}{{Baume}, G.} \& \bibinfo{author}{{Feinstein}, C.}
\newblock \bibinfo{journal}{\bibinfo{title}{{Binary and grouped open clusters: A new catalogue}}}.
\newblock {\emph{\JournalTitle{\aap}}} \textbf{\bibinfo{volume}{693}}, \bibinfo{pages}{A218}, \doiprefix\url{10.1051/0004-6361/202452672} (\bibinfo{year}{2025}).
\newblock \eprint{2412.05376}.

\bibitem{Hunt2023}
\bibinfo{author}{{Hunt}, E.~L.} \& \bibinfo{author}{{Reffert}, S.}
\newblock \bibinfo{journal}{\bibinfo{title}{{Improving the open cluster census. II. An all-sky cluster catalogue with Gaia DR3}}}.
\newblock {\emph{\JournalTitle{\aap}}} \textbf{\bibinfo{volume}{673}}, \bibinfo{pages}{A114}, \doiprefix\url{10.1051/0004-6361/202346285} (\bibinfo{year}{2023}).
\newblock \eprint{2303.13424}.

\bibitem{hunt2024improving}
\bibinfo{author}{Hunt, E.~L.} \& \bibinfo{author}{Reffert, S.}
\newblock \bibinfo{journal}{\bibinfo{title}{Improving the open cluster census-iii. using cluster masses, radii, and dynamics to create a cleaned open cluster catalogue}}.
\newblock {\emph{\JournalTitle{Astronomy \& Astrophysics}}} \textbf{\bibinfo{volume}{686}}, \bibinfo{pages}{A42} (\bibinfo{year}{2024}).

\bibitem{King1962}
\bibinfo{author}{{King}, I.}
\newblock \bibinfo{journal}{\bibinfo{title}{{The structure of star clusters. I. an empirical density law}}}.
\newblock {\emph{\JournalTitle{\aj}}} \textbf{\bibinfo{volume}{67}}, \bibinfo{pages}{471}, \doiprefix\url{10.1086/108756} (\bibinfo{year}{1962}).

\bibitem{Rangwal2019}
\bibinfo{author}{{Rangwal}, G.}, \bibinfo{author}{{Yadav}, R.~K.~S.}, \bibinfo{author}{{Durgapal}, A.}, \bibinfo{author}{{Bisht}, D.} \& \bibinfo{author}{{Nardiello}, D.}
\newblock \bibinfo{journal}{\bibinfo{title}{{Astrometric and photometric study of NGC 6067, NGC 2506, and IC 4651 open clusters based on wide-field ground and Gaia DR2 data}}}.
\newblock {\emph{\JournalTitle{\mnras}}} \textbf{\bibinfo{volume}{490}}, \bibinfo{pages}{1383--1396}, \doiprefix\url{10.1093/mnras/stz2642} (\bibinfo{year}{2019}).
\newblock \eprint{1909.08810}.

\bibitem{Krone-Martins2014}
\bibinfo{author}{{Krone-Martins}, A.} \& \bibinfo{author}{{Moitinho}, A.}
\newblock \bibinfo{journal}{\bibinfo{title}{{UPMASK: unsupervised photometric membership assignment in stellar clusters}}}.
\newblock {\emph{\JournalTitle{\aap}}} \textbf{\bibinfo{volume}{561}}, \bibinfo{pages}{A57}, \doiprefix\url{10.1051/0004-6361/201321143} (\bibinfo{year}{2014}).
\newblock \eprint{1309.4471}.

\bibitem{Campello2013}
\bibinfo{author}{Campello, R. J. G.~B.}, \bibinfo{author}{Moulavi, D.} \& \bibinfo{author}{Sander, J.}
\newblock \bibinfo{title}{Density-based clustering based on hierarchical density estimates}.
\newblock In \bibinfo{editor}{Pei, J.}, \bibinfo{editor}{Tseng, V.~S.}, \bibinfo{editor}{Cao, L.}, \bibinfo{editor}{Motoda, H.} \& \bibinfo{editor}{Xu, G.} (eds.) \emph{\bibinfo{booktitle}{Advances in Knowledge Discovery and Data Mining}}, \bibinfo{pages}{160--172} (\bibinfo{publisher}{Springer Berlin Heidelberg}, \bibinfo{address}{Berlin, Heidelberg}, \bibinfo{year}{2013}).

\bibitem{McInnes2017}
\bibinfo{author}{{McInnes}, L.}, \bibinfo{author}{{Healy}, J.} \& \bibinfo{author}{{Astels}, S.}
\newblock \bibinfo{journal}{\bibinfo{title}{{hdbscan: Hierarchical density based clustering}}}.
\newblock {\emph{\JournalTitle{The Journal of Open Source Software}}} \textbf{\bibinfo{volume}{2}}, \bibinfo{pages}{205}, \doiprefix\url{10.21105/joss.00205} (\bibinfo{year}{2017}).

\bibitem{Tarricq2022}
\bibinfo{author}{{Tarricq}, Y.} \emph{et~al.}
\newblock \bibinfo{journal}{\bibinfo{title}{{Structural parameters of 389 local open clusters}}}.
\newblock {\emph{\JournalTitle{\aap}}} \textbf{\bibinfo{volume}{659}}, \bibinfo{pages}{A59}, \doiprefix\url{10.1051/0004-6361/202142186} (\bibinfo{year}{2022}).
\newblock \eprint{2111.05291}.

\bibitem{Zhong2022}
\bibinfo{author}{{Zhong}, J.}, \bibinfo{author}{{Chen}, L.}, \bibinfo{author}{{Jiang}, Y.}, \bibinfo{author}{{Qin}, S.} \& \bibinfo{author}{{Hou}, J.}
\newblock \bibinfo{journal}{\bibinfo{title}{{New Insights into the Structure of Open Clusters in the Gaia Era}}}.
\newblock {\emph{\JournalTitle{\aj}}} \textbf{\bibinfo{volume}{164}}, \bibinfo{pages}{54}, \doiprefix\url{10.3847/1538-3881/ac77fa} (\bibinfo{year}{2022}).
\newblock \eprint{2206.04904}.

\bibitem{Gao2020}
\bibinfo{author}{{Gao}, X.}
\newblock \bibinfo{journal}{\bibinfo{title}{{5D memberships and fundamental properties of the old open cluster NGC 6791 based on Gaia-DR2}}}.
\newblock {\emph{\JournalTitle{\apss}}} \textbf{\bibinfo{volume}{365}}, \bibinfo{pages}{24}, \doiprefix\url{10.1007/s10509-020-3738-2} (\bibinfo{year}{2020}).

\bibitem{Nasser2025d}
\bibinfo{author}{{Ahmed}, N.~M.}
\newblock \bibinfo{journal}{\bibinfo{title}{{A comprehensive study of the old open cluster NGC 6791 using Gaia DR3 data and BV photometry}}}.
\newblock {\emph{\JournalTitle{\mnras}}} \textbf{\bibinfo{volume}{543}}, \bibinfo{pages}{1584--1601}, \doiprefix\url{10.1093/mnras/staf1469} (\bibinfo{year}{2025}).

\bibitem{Lindegren2021}
\bibinfo{author}{{Lindegren}, L.} \emph{et~al.}
\newblock \bibinfo{journal}{\bibinfo{title}{{Gaia Early Data Release 3. Parallax bias versus magnitude, colour, and position}}}.
\newblock {\emph{\JournalTitle{\aap}}} \textbf{\bibinfo{volume}{649}}, \bibinfo{pages}{A4}, \doiprefix\url{10.1051/0004-6361/202039653} (\bibinfo{year}{2021}).
\newblock \eprint{2012.01742}.

\bibitem{Bailer-Jones2021}
\bibinfo{author}{{Bailer-Jones}, C.~A.~L.}, \bibinfo{author}{{Rybizki}, J.}, \bibinfo{author}{{Fouesneau}, M.}, \bibinfo{author}{{Demleitner}, M.} \& \bibinfo{author}{{Andrae}, R.}
\newblock \bibinfo{journal}{\bibinfo{title}{{Estimating Distances from Parallaxes. V. Geometric and Photogeometric Distances to 1.47 Billion Stars in Gaia Early Data Release 3}}}.
\newblock {\emph{\JournalTitle{\aj}}} \textbf{\bibinfo{volume}{161}}, \bibinfo{pages}{147}, \doiprefix\url{10.3847/1538-3881/abd806} (\bibinfo{year}{2021}).
\newblock \eprint{2012.05220}.

\bibitem{Marigo2017}
\bibinfo{author}{{Marigo}, P.} \emph{et~al.}
\newblock \bibinfo{journal}{\bibinfo{title}{{A New Generation of PARSEC-COLIBRI Stellar Isochrones Including the TP-AGB Phase}}}.
\newblock {\emph{\JournalTitle{\apj}}} \textbf{\bibinfo{volume}{835}}, \bibinfo{pages}{77}, \doiprefix\url{10.3847/1538-4357/835/1/77} (\bibinfo{year}{2017}).
\newblock \eprint{1701.08510}.

\bibitem{Spada2017}
\bibinfo{author}{{Spada}, F.}, \bibinfo{author}{{Demarque}, P.}, \bibinfo{author}{{Kim}, Y.~C.}, \bibinfo{author}{{Boyajian}, T.~S.} \& \bibinfo{author}{{Brewer}, J.~M.}
\newblock \bibinfo{journal}{\bibinfo{title}{{The Yale-Potsdam Stellar Isochrones}}}.
\newblock {\emph{\JournalTitle{\apj}}} \textbf{\bibinfo{volume}{838}}, \bibinfo{pages}{161}, \doiprefix\url{10.3847/1538-4357/aa661d} (\bibinfo{year}{2017}).
\newblock \eprint{1703.03975}.

\bibitem{Bressan2012}
\bibinfo{author}{{Bressan}, A.} \emph{et~al.}
\newblock \bibinfo{journal}{\bibinfo{title}{{PARSEC: stellar tracks and isochrones with the PAdova and TRieste Stellar Evolution Code}}}.
\newblock {\emph{\JournalTitle{\mnras}}} \textbf{\bibinfo{volume}{427}}, \bibinfo{pages}{127--145}, \doiprefix\url{10.1111/j.1365-2966.2012.21948.x} (\bibinfo{year}{2012}).
\newblock \eprint{1208.4498}.

\bibitem{Nasser2025a}
\bibinfo{author}{{Ahmed}, N.~M.} \& \bibinfo{author}{{Tadross}, A.~L.}
\newblock \bibinfo{journal}{\bibinfo{title}{{The photometry and kinematics studies of NGC 2509 derived from Gaia DR3}}}.
\newblock {\emph{\JournalTitle{Scientific Reports}}} \textbf{\bibinfo{volume}{15}}, \bibinfo{pages}{17676}, \doiprefix\url{10.1038/s41598-025-00383-x} (\bibinfo{year}{2025}).

\bibitem{Wang2019}
\bibinfo{author}{{Wang}, S.} \& \bibinfo{author}{{Chen}, X.}
\newblock \bibinfo{journal}{\bibinfo{title}{{The Optical to Mid-infrared Extinction Law Based on the APOGEE, Gaia DR2, Pan-STARRS1, SDSS, APASS, 2MASS, and WISE Surveys}}}.
\newblock {\emph{\JournalTitle{\apj}}} \textbf{\bibinfo{volume}{877}}, \bibinfo{pages}{116}, \doiprefix\url{10.3847/1538-4357/ab1c61} (\bibinfo{year}{2019}).
\newblock \eprint{1904.04575}.

\bibitem{Abdurro2022}
\bibinfo{author}{{Abdurro'uf}} \emph{et~al.}
\newblock \bibinfo{journal}{\bibinfo{title}{{The Seventeenth Data Release of the Sloan Digital Sky Surveys: Complete Release of MaNGA, MaStar, and APOGEE-2 Data}}}.
\newblock {\emph{\JournalTitle{\apjs}}} \textbf{\bibinfo{volume}{259}}, \bibinfo{pages}{35}, \doiprefix\url{10.3847/1538-4365/ac4414} (\bibinfo{year}{2022}).
\newblock \eprint{2112.02026}.

\bibitem{Nasser2024}
\bibinfo{author}{{Ahmed}, N.~M.}, \bibinfo{author}{{Bendary}, R.}, \bibinfo{author}{{Samir}, R.~M.} \& \bibinfo{author}{{Elhosseiny}, E.~G.}
\newblock \bibinfo{journal}{\bibinfo{title}{{A deep investigation of the poorly studied open cluster King 18 using CCD VRI, Gaia DR3 and 2MASS}}}.
\newblock {\emph{\JournalTitle{Scientific Reports}}} \textbf{\bibinfo{volume}{14}}, \bibinfo{pages}{23777}, \doiprefix\url{10.1038/s41598-024-70133-y} (\bibinfo{year}{2024}).

\bibitem{Nasser2025b}
\bibinfo{author}{{Ahmed}, N.~M.} \& \bibinfo{author}{{Darwish}, M.~S.}
\newblock \bibinfo{journal}{\bibinfo{title}{{Tidal tail identification and detailed analysis of the open star cluster King 13 using Gaia DR3 and 2MASS}}}.
\newblock {\emph{\JournalTitle{Scientific Reports}}} \textbf{\bibinfo{volume}{15}}, \bibinfo{pages}{18033}, \doiprefix\url{10.1038/s41598-025-96923-6} (\bibinfo{year}{2025}).

\bibitem{Richer2021}
\bibinfo{author}{{Richer}, H.~B.} \emph{et~al.}
\newblock \bibinfo{journal}{\bibinfo{title}{{Massive White Dwarfs in Young Star Clusters}}}.
\newblock {\emph{\JournalTitle{\apj}}} \textbf{\bibinfo{volume}{912}}, \bibinfo{pages}{165}, \doiprefix\url{10.3847/1538-4357/abdeb7} (\bibinfo{year}{2021}).
\newblock \eprint{2101.08300}.

\bibitem{SciPy2020}
\bibinfo{author}{Virtanen, P.} \emph{et~al.}
\newblock \bibinfo{journal}{\bibinfo{title}{{{SciPy} 1.0: Fundamental Algorithms for Scientific Computing in Python}}}.
\newblock {\emph{\JournalTitle{Nature Methods}}} \textbf{\bibinfo{volume}{17}}, \bibinfo{pages}{261--272}, \doiprefix\url{10.1038/s41592-019-0686-2} (\bibinfo{year}{2020}).

\bibitem{Salpeter1955}
\bibinfo{author}{{Salpeter}, E.~E.}
\newblock \bibinfo{journal}{\bibinfo{title}{{The Luminosity Function and Stellar Evolution.}}}
\newblock {\emph{\JournalTitle{\apj}}} \textbf{\bibinfo{volume}{121}}, \bibinfo{pages}{161}, \doiprefix\url{10.1086/145971} (\bibinfo{year}{1955}).

\bibitem{Almeida2023}
\bibinfo{author}{{Almeida}, A.}, \bibinfo{author}{{Monteiro}, H.} \& \bibinfo{author}{{Dias}, W.~S.}
\newblock \bibinfo{journal}{\bibinfo{title}{{Revisiting the mass of open clusters with Gaia data}}}.
\newblock {\emph{\JournalTitle{\mnras}}} \textbf{\bibinfo{volume}{525}}, \bibinfo{pages}{2315--2340}, \doiprefix\url{10.1093/mnras/stad2291} (\bibinfo{year}{2023}).
\newblock \eprint{2307.15182}.

\bibitem{Spitzer1987}
\bibinfo{author}{{Spitzer}, L.}
\newblock \emph{\bibinfo{title}{{Dynamical evolution of globular clusters}}} (\bibinfo{year}{1987}).

\bibitem{kupper2010tidal}
\bibinfo{author}{K{\"u}pper, A.~H.}, \bibinfo{author}{Kroupa, P.}, \bibinfo{author}{Baumgardt, H.} \& \bibinfo{author}{Heggie, D.~C.}
\newblock \bibinfo{journal}{\bibinfo{title}{Tidal tails of star clusters}}.
\newblock {\emph{\JournalTitle{Monthly Notices of the Royal Astronomical Society}}} \textbf{\bibinfo{volume}{401}}, \bibinfo{pages}{105--120} (\bibinfo{year}{2010}).

\end{thebibliography}
%



%
\end{document}